\documentclass{emulateapj}

\begin{document}

\shorttitle{HST imaging and Keck Spectroscopy of $I$-drops}
\shortauthors{E.~R.~Stanway et al.}

\title{HST imaging and Keck Spectroscopy of
 $z\approx 6$ $I$-band Drop-Out Galaxies in the {\em ACS} GOODS Fields}

\author{Elizabeth R.~Stanway, Andrew~J.~Bunker, Richard G.~McMahon}
\affil{Institute of Astronomy, University of Cambridge, Madingley Road, Cambridge, CB3\,0HA, U.K.\\ {\tt
email: ers, bunker, rgm@ast.cam.ac.uk}}
\author{Richard S.~Ellis, Tommaso Treu\footnote{Current Address: Department of Physics and Astronomy, University of California at Los Angeles, Los Angeles, CA 90095}\footnote{Hubble Fellow}}
\affil{California Institute of Technology, Mail Stop 105-24, Pasadena, CA~91109, U.S.A.\\ {\tt
email: rse, tt@astro.caltech.edu}}
\and
\author{Patrick J.~McCarthy}
\affil{Observatories of the Carnegie Institute of Washington, Santa Barbara Street, Pasadena, CA~91101, U.S.A.\\ {\tt
email: pmc2@ociw.edu}}

\begin{abstract} 
We measure the surface density of $i'$-band dropout galaxies at $z\sim
6$ through wide-field {\em HST/ACS} imaging and ultra-deep Keck/DEIMOS
spectroscopy. Using deep {\em HST/ACS} SDSS-$i'$ (F775W) and SDSS-$z'$
(F850LP) imaging from the Great Observatories Origins Deep Survey
North (GOODS-N; 200\,arcmin$^2$), we identify 9 $i'$-drops satisfying
$(i'-z')_{\rm AB}>1.5$ to a depth of $z'_{\rm AB}=25.6$ (corresponding
to $L^*_{UV}$ at $z\sim3$). We use $HK'$ imaging data to improve the
fidelity of our sample, discriminating against lower redshift red
galaxies and cool Galactic stars. Three $i'$-drops are consistent with
M/L/T dwarf stars. We present ultra-deep Keck/DEIMOS spectroscopy of
10 objects from our combined GOODS-N and GOODS-S $i'$-drop sample. We
detect Lyman-$\alpha$ emission at $z=5.83$ from one object in the
GOODS-S field, which lies only 8arcmin (i.e. $3\,h^{-1}_{70}$\,Mpc)
away from a previously confirmed $z=5.78$ object. One possible
Lyman-$\alpha$ emitter at $z=6.24$ is found in the GOODS-N field
(although identification of this spatially-offset emission line is
ambiguous). Using the rest-frame UV continuum from our 6 candidate
$z\sim6$ galaxies from the GOODS-N field, we determine a lower limit
to the unobscured volume-averaged global star formation rate at
$z\sim6$ of $(5.4\pm2.2)\times 10^{-4}\,h_{70}\,M_{\odot}\,{\rm
yr}^{-1}\,{\rm Mpc}^{-3}$.  We find that the cosmic star formation
density in Lyman Break galaxies (LBGs) with unobscured star formation
rates $>15\,M_{\odot}\,{\rm yr}^{-1}$ falls by a factor of 8 between
$z\sim3$ and $z\sim6$. Hence the luminosity function of LBGs must
evolve in this redshift interval: a constant integrated star formation
density at $z>3$ requires a much steeper faint-end slope, or a
brighter characteristic luminosity. This result is in agreement with
our previous measurement from the GOODS-S field, indicating that
cosmic variance is not a dominant source of uncertainty.
\end{abstract} 
\keywords{galaxies: formation -- galaxies: evolution -- galaxies: starburst -- 
galaxies: high redshift -- ultraviolet: galaxies -- surveys}

\maketitle

\section{Introduction}
\label{sec:intro}

In recent years the identification and study of very high redshift
galaxies ($z>5$) has been an area of active and rapidly-advancing
research.  The increasing availability of 8- and 10- meter class
telescopes, combined with the development of modern instrumentation such
as the Advanced Camera for Surveys ({\em ACS}) on the Hubble Space
Telescope ({\em HST}) has allowed imaging of ever fainter and more
distant galaxies.

While large ground-based telescopes 
have made possible spectroscopic confirmation of redshifts for galaxies
beyond $z\sim6$ (e.g. Hu et al.\ 2002), this process is expensive in
telescope time and is only possible for those objects with strong
emission lines or which are lensed by intervening objects (e.g. 
Ellis et al.\ 2001). As a result,
the use of ``photometric redshift'' selection from broad-band colors
(such as the Lyman-break technique of Steidel, Pettini \& Hamilton 1995)
has become a widespread method of identifying large samples of
high-redshift galaxies.

In Stanway, Bunker and McMahon (2003, hereafter paper 1) we extended the
Lyman-break selection technique to $z\sim 6$, and described the
photometric selection of $i'$-band drop-out candidate high redshift
objects in the Chandra Deep Field South (CDFS).
We demonstrated that public data from the Advanced Camera for Surveys
({\em ACS} -- Ford et al.\ 2002) on {\em HST}, released as part of the
Great Observatories Origins Deep Survey (GOODS -- Dickinson \&
Giavalisco 2003\footnote{see {\tt
http://www.stsci.edu/ftp/science/goods/}}), is of sufficient depth and
volume to detect very high redshift galaxies, and that color selection
can be used to reject less distant objects. We used this data set to
examine the space density of UV-luminous starburst galaxies at $z\approx
6$ in the GOODS-South field (the CDFS) and to place a lower limit on the
comoving global star formation rate at this epoch.  Furthermore, in
Bunker et al.\ (2003, hereafter paper 2) we presented deep spectroscopy
for object SBM03\#3, confirming its redshift as $z=5.78$ and
illustrating the effectiveness of our $i'$-drop selection.

In Section~\ref{sec:phot} of this paper we present
a similar selection of 9 $i'$-drop objects with
large $(i'-z')_{\rm AB}$ colors in the GOODS/{\em ACS} northern field
(hereafter GOODS-N), centered on the Hubble Deep Field North (HDFN,
Williams et al.\ 1996). Deep
spectroscopy using the DEIMOS spectrograph on the Keck\,\mbox{\sc{II}}
telescope is presented in Section~\ref{sec:spec} for half of our
$(i'-z')_{\rm AB}>1.5$ sample, including 5 $i'$-drops in the
GOODS-N. We also revisit our GOODS-S $i'$-drop sample from paper
1, and present 4 spectra in this field (in addition to the $z=5.78$ GOODS-S
galaxy reported in Paper~2).

In Section~\ref{sec:discussion}, we discuss the effects of completeness
corrections and surface brightness effects on our sample.
In Section~\ref{sec:sfr} we estimate the global star
formation rate at $z\sim$6, and
contrast the GOODS-N dataset with that
for the GOODS-S, enabling us to address the issue of cosmic variance. We
also discuss the effects of completeness corrections, dust extinction
and evolution of the Lyman-break galaxy luminosity function, and we
compare our results to other recent work in this field.  Our conclusions
are presented in Section~\ref{sec:conclusion}.

For consistency with our earlier work, we adopt the following
cosmology: a $\Lambda$-dominated, flat Universe with
$\Omega_{\Lambda}=0.7$, $\Omega_{M}=0.3$ and $H_{0}=70\,h_{70} {\rm
km\,s}^{-1}\,{\rm Mpc}^{-1}$. All magnitudes in this paper are quoted
in the AB system (Oke \& Gunn 1983) and the Madau (1995) prescription,
extended to $z=6$, is used where necessary to estimate absorption due
to the intergalactic medium.

\section{GOODS-North $i$-Drops: Photometry and Candidate Selection}
\label{sec:phot}

\subsection{GOODS {\em HST/ACS} imaging and $i'$-Drop Selection}

We present here a high redshift candidate selection based on the first
four epochs of GOODS {\em HST/ACS} observations of the GOODS-N available at
the time of writing.
Each epoch of data comprises imaging of either 15 or 16 adjacent
`tiles' in each of the F606W ($V$, 0.5 orbits), F775W (SDSS-$i'$, 0.5
orbits) and F850LP (SDSS-$z'$, 1.0 orbits) broad-band filters. Three
orbits in the F435W $B$-filter 
were obtained in a single epoch at the start of the observing campaign.
As in paper 1, we analyze v0.5 of the publically-released reduced
data\footnote{available from {\tt
ftp://archive.stsci.edu/pub/hlsp/goods/}}.

We use the photometric calibration determined by the GOODS team:
\begin{displaymath}
\mathrm{mag}_{AB} = \mathrm{zeropoint}-2.5\log_{10}(\mathrm{Count~rate} 
/ {\mathrm s}^{-1}), 
\end{displaymath}
and note that the zeropoints for the {\em ACS} instrument changed
between epochs 1 and 2 of the GOODS-N observations.  In epoch 1 $AB$
magnitude zeropoints for the $v$-band (F606W), $i'$-band (F775W) and
$z'$-band (F850LP) are $26.505$, $25.656$ and $24.916$ respectively. In
Epoch 2 and thereafter the zeropoints for the $v$-band, $i'$-band and
$z'$-band are altered to $26.493$, $25.641$ and $24.843$
respectively. The $B$-band (F439W) zeropoint is $25.662$.

We have corrected for the small amount of foreground Galactic extinction
toward the HDFN using the {\it COBE}/DIRBE \& {\it IRAS}/ISSA dust maps
of Schlegel, Finkbeiner \& Davis (1998). For the GOODS-N field, Galactic 
extinction is given by E(B-V) = 0.012 mag, equivalent to
$A_{\mathrm F775W}=0.025$ and $A_{\mathrm F850LP}=0.018$.

For consistency, given minor discrepancies between object coordinates in
different epochs, the astrometry presented in this paper is taken from
the world coordinate system provided on the third epoch of the GOODS v0.5
released data.  It should be noted that the first epoch of data was
released with only an approximate astrometry solution although this was
corrected subsequently.

Within each epoch of data, object detection 
was performed using the SExtractor photometry software package (Bertin
\& Arnouts 1996). Fixed circular apertures $1\farcs0$ in diameter were
trained in the $z'$-band image and the identified apertures were used to
measure the flux at the same spatial location in the $i'$-band image,
running SExtractor in `two-image' mode. This allows for identification
of any object securely detected in the $z'$-band but faint or undetected
at shorter wavelengths -- the expected signature of high redshift
objects.  For object identification, we demanded at least 5 adjacent
pixels above a flux threshold of $2\sigma$ per pixel ($0.01\,{\rm
counts\,pixel}^{-1}\,{\rm s}^{-1}$) on the drizzled data (with a pixel
scale of 0\farcs05~pixel$^{-1}$).

A preliminary catalog of detected objects was created for each epoch of
data. In the compilation of the catalogues a number of `figure-eight'
shaped optical reflections and diffraction spikes near bright stars were
masked, as was the gap between the two {\em ACS} CCDs and its vicinity in each
tile. Excluding the masked regions, the total survey area in the GOODS-N is
200 \,arcmin$^{2}$, although approximately 20\% of this area is only
observed in half of the available epochs.  We estimate that the fraction
of each image contaminated by bright foreground objects is small ($<2$
per cent).

To select $z\sim 6$ galaxies, we adapt the now widely-applied Lyman
break technique pioneered at $z\sim 3$ by Steidel and co-workers
(Steidel et al.\ 1995, 1996). At $z\sim 6$ the integrated optical
depth of the Lyman-$\alpha$ forest is $\gg 1$ and hence the continuum
break in an object's spectrum at the wavelength of Lyman-$\alpha$ is
large, and can be identified using just two filters: one in the Lyman
forest region, where there is essentially no flux transmitted, and a
second longward of the Lyman-$\alpha$ line ($\rm
\lambda_{rest}=1216$\,\AA). At $z\sim 6$ Lyman-$\alpha$ emission lies
at an observed wavelength of $\rm \lambda_{obs}\sim 8500$\,\AA , in
the SDSS-$z'$ (F850LP) band, and we use the SDSS-$i'$ (F775W) band to
cover the Lyman forest wavelengths. As discussed in paper~1, the sharp
band-pass edges of these SDSS filters aid in the clean selection of
high-redshift objects (such as the $z\sim 6$ QSOs of Fan et al.\
2001), and in paper~2 we spectroscopically confirmed an $i'$-drop
selected by this technique to be a star-forming galaxy at $z=5.78$.

To quantify this, simulations were performed in which an ensemble of
galaxies with a distribution of spectral slopes and luminosities
similar to those of the well-studied Lyman-break population at
$z\sim3$ (Adelberger \& Steidel 2000) were generated.  These were
uniformly distributed in the redshift range $4<z<7$ and the effects of
absorption in the IGM simulated using the Madau (1995)
prescription. The $i'-z'$ colours of these galaxies were then
calculated and hence the selection completeness in each redshift bin
assessed. In these simulations, $>90$\% of galaxies brighter than our
magnitude limit at $z>5.6$ and $>99$\% of galaxies at $z>5.7$
satisfied our colour cut criterion of $(i'-z')_{\rm AB}>1.5$. The
fraction of $z>4$ interlopers with these colours was
negligible($<1$\%).  The luminosity function and distribution in
spectral slope of galaxies at $z\sim6$ are currently unknown (see
section \ref{sec:lum}) so these simulations are indicative only but do
suggest that this technique would successfully recover a population
equivalent to that at $z\sim3$.

We impose a magnitude limit corresponding to $L^*$ in the rest-frame
UV for the Lyman-break population at $z=3$ (Steidel et al.\ 1999):
$M_{AB}^{*}(1500{\rm \AA })=-21.1$\,mag, which translates to $z'_{\rm
AB}=25.6$\,mag at $z=5.8$ where our selection sensitivity peaks (see
section \ref{sec:volume}) and to an unobscured star formation rate of
$15\,M_{\odot}\,{\rm yr}^{-1}$ (see section \ref{sec:sfr}).  This
cut, a $7\,\sigma$ detection in each epoch of the GOODS/{\em ACS}
data, ensures that each object is securely detected in $z'$.

Randomly placed artificial galaxies were generated in each epoch,
using the {\sc IRAF.Artdata} package, with $z'$ magnitudes in the
range 20-28 and distributed acording to a Schecter luminosity function
with a faint end slope $\alpha=-1.6$ and $m^*=25.6$, the parameters of
the Lyman break population at $z=3$ projected to $z=6$. We confirm
that at our magnitude cut and surface brightness limit (see Figure
\ref{fig:zfwhm}), we reliably recover $>$98 per cent of such galaxies
in the $z'$ band in each epoch, with the remaining 2 per cent lost to
crowding (confusion due to blended objects).

\begin{figure}
\plotone{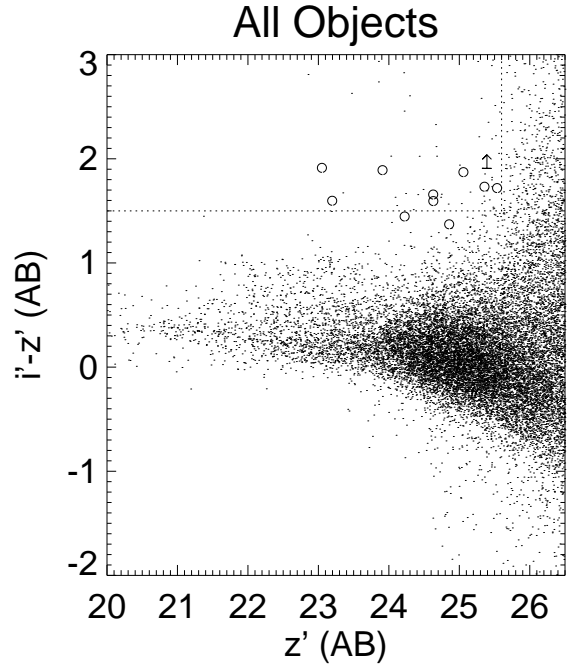}
\caption{Color selection of our candidates showing the $(i'-z')_{\rm
AB}>1.5$,\ $z'_{\rm AB}<25.6$\,mag selection criteria applied in each
epoch and indicating the colors of the candidate objects in table
\ref{tab}. A further two objects lying within $1\sigma$ our $i'-z'$
cut are included for completeness. Note that the catalog presented
here is not clean and indicates the large number of data artifacts
that can mimic these $i'-z'$ colours. A number of `objects' identified
in each epoch were in fact connected to unmasked satellite tracks or
optical refraction artifacts and these objects were rejected from the
final selections.}
\label{plot:colour}
\end{figure}

In order to minimize the effects of photometric errors, un-rejected
cosmic rays and transient objects on our final candidate list, an
independent selection of objects satisfying the criteria of
$(i'-z')_{\rm AB}>1.5$ and $z'_{\rm AB}<25.6$\,mag was made on each
epoch of the GOODS data set (see Figure\,\ref{plot:colour}). This
colour cut criterion allows seperation of low from high redshift
galaxies (subject to photometric scatter) as shown by figure 2 in
paper 1 while minimising the contamination due to elliptical galaxies
at $z\sim2$. In order to ensure the completeness of the final sample,
objects detected with magnitudes and colors within approximately
2$\sigma$ of our final color and magnitude cut in each epoch were also
considered. The objects in the resulting sub-catalog were then
examined for time-varying transient behavior across those epochs in
which they were observed, and a number of objects which satisfied our
color criterion in individual epochs were rejected at this stage.

The fluxes of the remaining candidate objects were then measured
directly within a 1\arcsec\ diameter aperture in each epoch and these
fluxes averaged to give a final magnitude, thereby reducing the
photometric error for each object. To convert to total magnitudes, an
aperture correction of 0.09\,mag was applied to all objects,
determined from aperture photometry of bright but unsaturated point
sources. Although this correction will underestimate the magnitude of
very extended sources, the candidate objects presented in this paper
are all compact and it has not been necesary to apply different
aperture corrections to different candidates.

Although a deeper candidate selection may have been possible using
coadded images for selection, we have found that the advantages of
being able to eliminate spurious and transient sources by epoch-epoch
comparison are significant. An ideal algorithm for co-adding the
images would reject all deviant pixels allowing identification of
fainter objects, however such an algorithm is in practise difficult to
obtain with just three or even five epochs of data and coadding the
data here will lead to the loss of time-dependent information or
information on poorly rejected spurious pixels.  In several cases the
time-averaged colour of an object was significantly affected by hot
pixels or poorly rejected cosmic rays in a single epoch. The selection
method outlined above allowed epochs for which the data were
unreliable to be rejected when making a final selection. As a result
we have chosen to work at a relatively bright magnitude limit on the
shallower single epoch data. the completeness of our resulting sample
is discused in section \ref{sec:discussion}.

\begin{figure}
\plotone{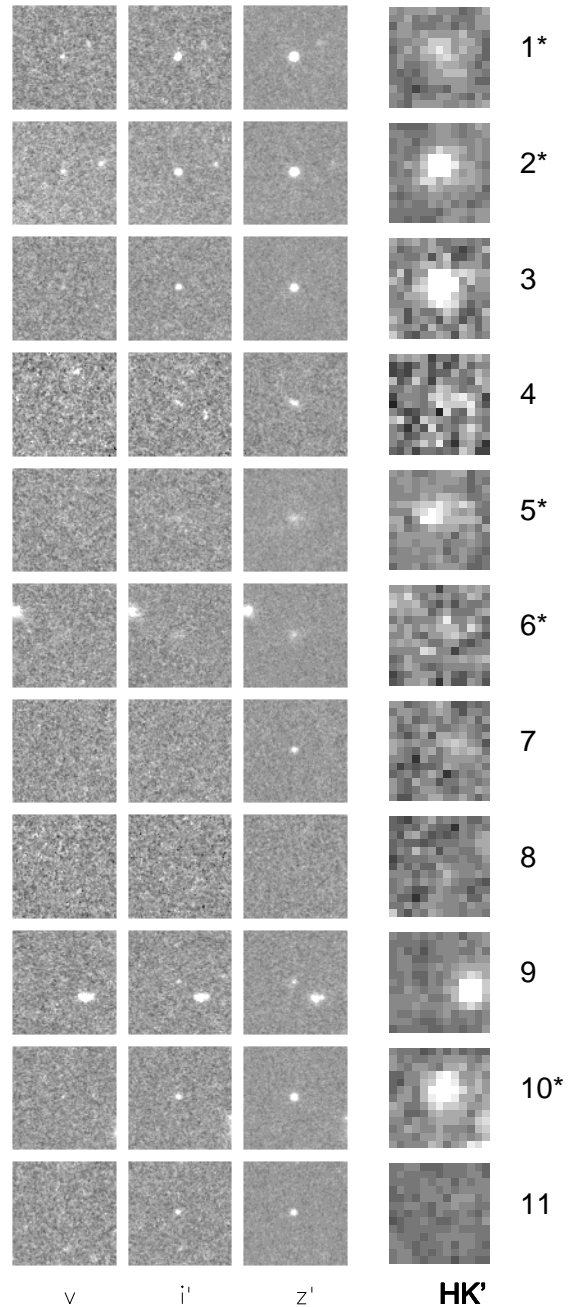}
\caption{$i'$-drops in the GOODS-N field: {\em HST/ACS} and Hawaii
$HK'$, 4 arcsecond boxes. {\em ACS} images are constructed from all
available epochs. An asterisk indicates that an object forms part of our
spectroscopic sample.}
\label{fig:images}
\end{figure}

\begin{figure}
\plotone{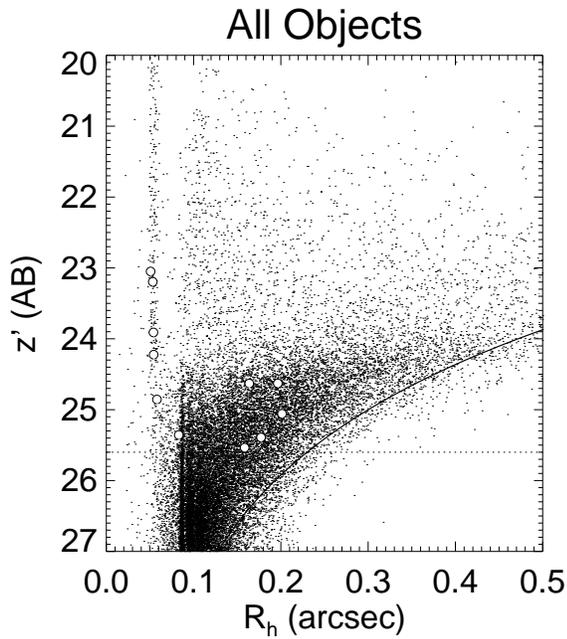}
\caption{The magnitude--size distribution of our GOODS-N $i$-drops
(circles) and objects in the field. Note the cut-off at large half-light
radii (R$_h$) due to surface brightness dimming (the solid curve is our
completeness limit). Our magnitude limit is marked with a dotted line.}
\label{fig:zfwhm}
\end{figure}

In table \ref{tab} we present details of the nine objects meeting our
selection criteria. In addition the two objects which lie within $1\sigma$ of
being selected are included in the following discussions for completeness. 
Postage stamp images of each object in the $v$, $i'$, $z'$ and $HK'$ bands 
(see section \ref{sec:ir}) are presented in figure \ref{fig:images}. None of 
the $i'$-drops in table \ref{tab} were formally detected at
$>3\sigma$ in the F606W ($v$) or F435W ($B$) band ($v_{AB}<27.3$,
$B_{AB}<27.0$\,mag) of GOODS/{\em ACS} observations except candidate 6 which
may be subject to contamination from a nearby object
(figure~\ref{fig:images}).

As illustrated by figure \ref{fig:zfwhm}, 3 of the 9 $i'$-drop objects
in table \ref{tab} are unresolved in the GOODS/{\em ACS} $z'$ images (R$_h$
=0\farcs05). These objects may reasonably be Galactic stars, high
redshift QSOs or compact galaxies.  Previous studies have shown that
galaxies at very high redshift are barely resolvable with {\em HST} (and
are sometimes unresolved), examples including the $z>5$ candidates of
Bremer et al.\ (2003) which have R$_h\sim$0\farcs1-0\farcs3, and the
barely-resolved spectroscopically confirmed z=5.78 galaxy SBM03\#3 which
has an R$_h=$0\farcs08 (Paper 2). With this in mind, we consider each
object individually rather than requiring that all galaxy candidates are
resolved.

\subsection{X-ray properties: Non-Detection of $i'$-Drops by {\em Chandra}}
\label{sec:xray}

To avoid possible contamination of our sample by high redshift AGN we
examined publically-available deep data from the Chandra X-ray
telescope.

All of our candidates lie in the region surveyed in a 2\,Ms exposure
by the {\em Chandra} X-ray satellite (Alexander et al.\ 2003) and none
of these objects were detected by that survey, allowing us to place a
limit on their X-ray fluxes $0.5-2$\,keV (soft) and $2-8$\,keV (hard)
bands at between $1.9\times10^{-17}$ and $9.3\times10^{-17}$ ergs
cm$^{-2}$ s$^{-1}$ and between $1.1\times10^{-16}$ to
$7.5\times10^{-16}$ ergs cm$^{-2}$ s$^{-1}$ respectively, varying
according to the non-uniform exposure across the GOODS-N field
(D.M. Alexander, private communication).  A powerful AGN (quasar) is
expected to have a rest-frame 2-8 keV luminosity
$>10^{44}\,h^{-2}_{70}\,{\rm erg\,s}^{-1}$ (Barger et al.\ 2003),
which yields a flux at $z=5.8$ of $\sim3\times10^{-16}\,{\rm
erg\,s}^{-1}\,{\rm cm}^{-2}$. Thus we would expect to detect X-ray
luminous quasars for most of our sources at the relevant search
redshifts probed in this paper. 

This renders the identification of any of our candidates as powerful
high-redshift AGN unlikely although non-detection of x-ray flux at
these limits does not rule out the presence of fainter high redshift
AGN such as those observed in $\approx3\%$ of Lyman break galaxies at
$z\sim3$ (Steidel et al. 2002). In this $z\sim3$ sample, AGN
contribute less than 2\% of the total rest-frame UV
luminosity. Calculations in section \ref{sec:sfr} require an implicit
assumption is that the rest-frame flux is dominated by star
formation. Although an absence of luminous AGN does not necesarily
imply a UV continuum dominated by star formation, it does decrease the
likelihood of significantly non-stellar origins for the flux.

\begin{deluxetable*}{cccccccccccc}
\tablecaption{$i'$-drops in GOODS-N}
\tablewidth{0pt}
\tablehead{
ID & RA & Declination  &  $z'_{\rm AB}$  & $i'_{\rm AB}$ & $(i'-z')_{\rm AB}$   &$(z'-HK')_{\rm AB}$ & R$_h$ & SFR$^{z=5.8}_{\rm UV}$ & Final\\
 & (J2000) & (J2000) & & & & & arcsec &$h^{-2}_{70}\,M_{\odot}\,{\rm yr}^{-1}$ & Cut? }
\startdata
$ 1$\tablenotemark{$\ast$} & 12$^{\rm h}$36$^{\rm m}$58\fs13 & +62\degr18\arcmin51\farcs5 &  $ 23.05 \pm 0.02 $ & $  24.96 \pm 0.07 $ & $ 1.91 \pm 0.07$  & $ 0.21 \pm 0.11  $ &  $ 0.051$\tablenotemark{\dag}  & [140]     & No\\
$ 2$\tablenotemark{$\ast$} & 12$^{\rm h}$37$^{\rm m}$34\fs04 & +62\degr15\arcmin53\farcs4 &  $ 23.20 \pm 0.02 $ & $  24.79 \pm 0.06 $ & $ 1.60 \pm 0.06$  & $ 1.62 \pm  0.04 $ &  $ 0.053$\tablenotemark{\dag}  & [122]     & No\\
$ 3 $      & 12$^{\rm h}$36$^{\rm m}$38\fs85 & +62\degr14\arcmin51\farcs9 &  $ 23.91 \pm 0.04 $ & $  25.80 \pm 0.16 $ & $ 1.89 \pm 0.17$  & $ 1.89 \pm  0.05 $ &  $ 0.054$\tablenotemark{\dag}  & [63]      & No \\
$ 4 $      & 12$^{\rm h}$35$^{\rm m}$37\fs21 & +62\degr12\arcmin03\farcs5 &  $ 24.63 \pm 0.07 $ & $  26.13 \pm 0.25 $ & $ 1.59 \pm 0.26$  & $ <2.22     $ $(2\,\sigma)$     &  $ 0.164 $      & 33        & Yes\\
$ 5$\tablenotemark{$\ast$} & 12$^{\rm h}$36$^{\rm m}$48\fs47 & +62\degr19\arcmin02\farcs0 &  $24.63 \pm 0.07$   & $  26.29 \pm 0.27 $ & $  1.67 \pm 0.28$ & $1.70 \pm 0.08 $   &  $ 0.197 $       & 33        & Yes \\
$ 6$\tablenotemark{$\ast$} & 12$^{\rm h}$37$^{\rm m}$39\fs29 & +62\degr18\arcmin40\farcs2 &  $ 25.06 \pm 0.10 $ & $  26.93 \pm 0.54 $ & $ 1.87 \pm 0.55$  & $ <2.65       $  $(2\,\sigma)$   &  $ 0.201 $      & 22        & Yes\\
$ 7 $      & 12$^{\rm h}$35$^{\rm m}$52\fs35 & +62\degr11\arcmin42\farcs1 &  $ 25.36 \pm 0.14 $ & $  27.09 \pm 0.66 $ & $ 1.73 \pm 0.67$  & $ <2.95       $ $(2\,\sigma)$    &  $ 0.083 $      & 17        & Yes\\
$ 8 $      & 12$^{\rm h}$36$^{\rm m}$50\fs41 & +62\degr20\arcmin12\farcs6  &  $25.39  \pm 0.14$  & $ < 27.19          $ $(2\,\sigma)$ & $  > 1.71       $ $(2\,\sigma)$ & $<2.98     $   $(2\,\sigma)$  &  $ 0.177$        & 16        & Yes \\
$ 9 $      & 12$^{\rm h}$36$^{\rm m}$48\fs74 & +62\degr12\arcmin17\farcs1 &  $ 25.54 \pm 0.16 $ & $  < 27.19  $ $(2\,\sigma)$ & $ >1.56 $  $(2\,\sigma)$              & $< 1.93$\tablenotemark{$\ast\ast$}\ $(2\,\sigma)$& $ 0.159 $      & 14        & Yes \\
\tableline
$ 10$\tablenotemark{$\ast$} & 12$^{\rm h}$36$^{\rm m}$53\fs79 & +62\degr11\arcmin18\farcs2 &  $ 24.22 \pm 0.05 $ & $  25.67 \pm 0.14 $ & $ 1.45 \pm 0.15$  & $ 1.96  \pm 0.07 $ &  $ 0.054$\tablenotemark{\dag}  & [47]      & No \\
$ 11 $     & 12$^{\rm h}$36$^{\rm m}$49\fs79 & +62\degr16\arcmin24\farcs9 &  $ 24.86 \pm 0.09 $ & $  26.23 \pm 0.25 $ & $ 1.37 \pm 0.26$  & $< 2.45      $ $(2\,\sigma)$     &  $ 0.058$\tablenotemark{\dag}  & [26]      & No \\
\enddata
\tablenotetext{$\ast$}{ Keck/DEIMOS spectrum obtained}
\tablenotetext{\dag}{ Unresolved point source}
\tablenotetext{$\ast\ast$} { Additional Near-IR photometry is available for candidate 9 from the deep HST/NICMOS imaging of the HDFN (Thompson et al.\ 1999).This object has $F110W_{\rm AB} (J) = 25.10$ and $F160W_{\rm AB} (H) =24.60$ in this dataset.}
\tablecomments{Objects 10
and 11 (below the horizontal line) lie marginally outside our selection
criteria but are included for completeness.  
All magnitudes are measured 
within a 1\,arcsec diameter aperture.
The half-light radius, $R_h$, is defined as the
radius enclosing half the flux in the $z'$ band. 
The nominal star formation rate (SFR) is
calculated assuming objects are placed in the middle of our effective
volume (a luminosity-weighted redshift of $z =5.8$) and are shown in
square brackets if the candidate is not included in our final high
redshift selection.  \label{tab} }
\end{deluxetable*}

\subsection{Near-infrared Imaging}
\label{sec:ir}

\begin{figure}
\plotone{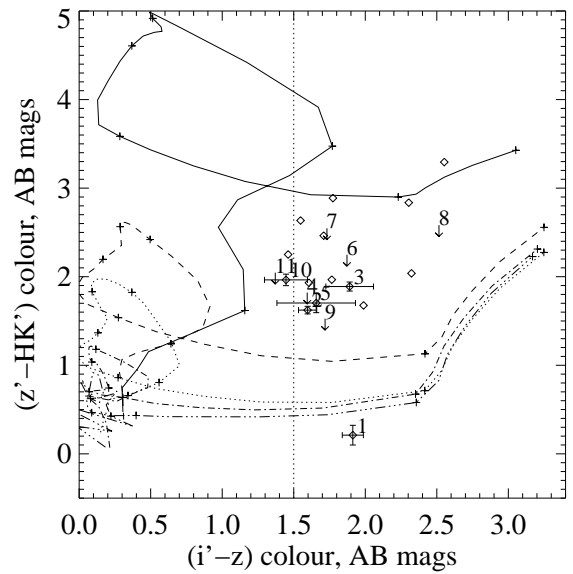}
\caption{Evolutionary tracks and color discrimination: the solid black
line indicates the evolutionary path followed by elliptical galaxies
with redshift (calculated using the Coleman, Wu and Weedman (1980)
spectral template). Also indicated by lines are evolutionary paths for
Sab (dashed), Sbc (dotted) and Im (dot-dash) galaxies and for a Kinney
et al.\ starburst model (dot-dot-dash). Tick marks on tracks indicate
unit redshift intervals between $z=0$ (lower left) and $z=7$ (upper
right). The vertical dotted line indicates our basic color-cut criterion
of $(i'-z')_{\rm AB}>1.5$. The colors of candidates in table \ref{tab}
are indicated with error bars or as upper limits in $(z'-HK')_{\rm
AB}$. The colors of typical L class stars are shown as small
diamonds, derived from the spectra from Reid et al.\ (2001). }
\label{plot:tracks}
\end{figure}

As figure \ref{plot:tracks} illustrates, a color cut of $(i'-z')_{\rm
AB}>1.5$ can be used to select galaxies with $z>5.6$.  
This color-cut criterion alone may also select 
elliptical galaxies at $z\simeq 2$
(e.g., Cimatti et al.\ 2002) and cool low-mass M/L/T stars (e.g.,
Kirkpatrick et al.\ 1999, Hawley et al.\ 2002).
We can guard against such contaminants by utilizing near-infrared photometry.
In particular the relatively blue $(z'-HK')>3$ color expected of
$z\simeq2$ ellipticals allows these objects to be distinguished from high redshift candidate objects.

In order to improve our discrimination between high and low redshift
objects we have supplemented the GOODS {\em
HST/ACS} data with publically-available near-infrared images from the
Hawaii-HDFN project\footnote{available from {\tt
http://www.ifa.hawaii.edu/$\sim$capak/hdf/}}. The $HK'$-band data is
described in Capak et al.\ (2003), and was obtained on the University of
Hawaii 2.2m telescope, with a pixel scale of $0\farcs3\,{\rm
pixel}^{-1}$. It covers an area of 0.11\,deg$^2$ to a depth of
$HK'_{AB}=22.5$ (encompassing the entire GOODS area), with the central
$9'\times9'$ region imaged to a greater depth of $HK'_{AB}=23.7$
($2\,\sigma$, 1\arcsec\ diameter aperture).

SExtractor was again used to identify objects in the $HK'$ images with 5
or more adjacent pixels exceeding the background noise by 2$\sigma$.
Matches for the $i'$-drop candidate objects were
assigned if positions of the sources were within $1\arcsec$, although we
carefully examined the $z'$-band image to see if nearby non-$i'$-drop
sources may instead be responsible for any near-infrared flux. An aperture 
correction of 0.4 mag was applied to the $1\arcsec$ aperture magnitudes
in order to estimate the total flux.

Low redshift $z\simeq1-2$ elliptical galaxies are known to be a
contaminant in colour selected samples at $z>5$ (Capak et
al.\ 2003, Bremer et al.\ 2003). Our $i'-z'$ colour cut criterion was
designed to exclude these but contamination is still possible due to
scatter in the photometric properties of these objects. Although the
near infrared colors of many of the objects presented in table
\ref{tab}, as shown on figure \ref{plot:tracks}, are presented only as
upper limits, those limits all place $(z'-HK')_{\rm AB}<2.5$ with the
exception of those for candidates 7 and 8. Given the red $(i'-z')_{\rm
AB}$ colour of candidate 8, only candidate 7 is marginally consistent
with being a low redshift elliptical galaxy. The lower limit of
$(z'-HK')_{\rm AB}<2.95$ for this object, however, does not preclude
the possibility that candidate 7 may also lie at much higher redshift.

The region of colour space occupied by L and M class stars is less
clearly distinct from that occupied by high redshift galaxies. Four
objects (candidates 2, 3, 5 and 10) have colors consistent with
identification as L class stars although all four are also still
consistant with a high redshift interpretation if the spectral slope
of high redshift objects is steeper than that observed in $z\sim3$
LBGs. Candidate 1 has colours which are consistant with either a high
redshift galaxy or a mid-M class star. Near-IR colours, together with
their spectroscopy and other properties, are considered in section
\ref{sec:fcands} where a final high redshift candidate selection is
made.

\section{$I$-Drops -- SPECTROSCOPIC PROPERTIES FOR THE HDF NORTH AND
CDF SOUTH SAMPLES}
\label{sec:spec}

\subsection{Spectroscopy - Observation Data}

We have obtained slit-mask spectra of about half our $i'$-drop sample,
using the Deep Imaging Multi-Object Spectrograph (DEIMOS, Faber et al.\
2002; Phillips et al.\ 2002) at the Cassegrain focus of the 10-m
Keck{\scriptsize~II} Telescope.  DEIMOS has 8 MIT/LL $2k\times 4k$ CCDs
with $15\,\mu$m pixels and an angular pixel scale of
0.1185\,arcsec\,pix$^{-1}$. The seeing was typically in the range
$0\farcs7-1\farcs0$ FWHM, smaller than or comparable to the slit width
of $1\farcs0$.

The observations were obtained using the Gold 1200\,line\,mm$^{-1}$
grating in first order.  The grating was tilted to place a central
wavelength of 8000 \AA\ on the detectors, and produce a dispersion of
$0.314$\,\AA\,pixel$^{-1}$. We sample a wavelength range of
approximately $\lambda\lambda_{\rm obs}\,6600-9100$\,\AA, corresponding
to a rest-frame wavelength of $\lambda\lambda_{\rm
rest}\,950-1300$\,\AA\ at $z\sim6$. Wavelength calibration was obtained
from Ne$+$Ar$+$Hg$+$Kr reference arc lamps. The spectral resolution was
measured to be $\Delta\lambda_{\rm FWHM}^{obs}\approx 1.4$\,\AA\
($\Delta v_{\rm FWHM}\approx 55$ km s$^{-1}$) from the sky lines. A
small 8\,\AA\ region in the middle of the wavelength range for each
object is unobserved as it falls in the gap between two CCDs. The
fraction of the spectrum affected by skylines is small ($<2$ per cent).
Flux calibration was determined using the spectra of the alignment stars
of known broad-band photometry ($I\approx 16-19$\,mag), used to position
the two slitmasks through $2''\times2''$ alignment boxes. 

In Paper~2 we already reported the confirmation of a $z=5.8$ galaxy
(SBM03\#3 in GOODS-S from Paper~1), based on a 5.5-hour DEIMOS
spectrum taken on UT 2003 January 08 \& 09. In
section~\ref{sec:CDFSspec} we present spectra from other GOODS-S
$i'$-drops on the same slitmask. Our spectroscopy reveals a second
$z=5.8$ Lyman-$\alpha$ emitter in the same field. The observations and
data reduction for this slitmask are as detailed in paper 2.

In addition, we have recently obtained ultra-deep spectroscopy of our
GOODS-N $i'$-drops over the 5 nights of U.T. 2003 April 2-6.  We used
several slitmasks to target a subset of five of the eleven objects in
table \ref{tab} (candidates 1, 2, 5, 6 and 10). The $i'$-drops were
placed on slitmasks for a primary program targetting $z\simeq 1$ E/S0
galaxies in the GOODS-N field (Treu et al.\ in prep). The selection of
$i'$-drops from table~\ref{tab} for the masks was purely geometric and
accordingly randomized.
The position angle of slits on the mask was chosen to be 45.3\degr\ so
as to align the large ($16.5\times5$\,arcmin) DEIMOS field of view with
the axis of the GOODS-N {\em HST/ACS} observations.  The spectra were
pipeline reduced using version 1.1.3 of the DEEP2 DEIMOS data reduction
software\footnote{available from {\tt
http://astron.berkeley.edu/$\sim$cooper/deep/spec2d/}}.

A total of 37.8\,ksec (10.5-hours) of on-source integration was obtained
in each of three slitmasks, and this was broken into individual
exposures each of duration 1800\,s. These spectra are therefore a factor
of $\sqrt{2}$ deeper than the 5.5-hour spectra of GOODS-S introduced in
paper~2. In 10.5 hours of DEIMOS spectroscopy we reach a flux limit of
$2\times10^{-18}\,{\rm ergs\,cm}^{-1}\,{\rm s}^{-1}$ ($5\,\sigma$) for
an emission line uncontaminated by sky lines and extracted over 8\AA\
(300\,km\,s$^{-1}$) and 1\arcsec .

The results of this spectroscopy are summarised in table \ref{tab2} and discussed in more detail in the following sections.

\begin{deluxetable}{ll}
\tablecaption{Summary of Spectroscopic Results\label{tab2}}
\tablewidth{0pt}
\tablehead{Object & Observation}
\startdata
GOODS-S SBM03\#1*  & Line emission at 8305\AA\ - identified as Lyman-$\alpha$\\
GOODS-S SBM03\#2  & No observed spectral features                           \\
GOODS-S SBM03\#3* & Line emission at 8245\AA\ - identified as Lyman-$\alpha$\\
GOODS-S SBM03\#5  & No observed spectral features                           \\
GOODS-S SBM03\#7* & No observed spectral features                           \\
GOODS-N\#1        & Continuum Flux longward of approx. 7200\AA\             \\
GOODS-N\#2        & Continuum Flux longward of approx. 7800\AA\             \\ 
GOODS-N\#5*       & No observed spectral features                           \\
GOODS-N\#6*       & Line emission at 8804\AA\ - identification uncertain    \\
GOODS-N\#10       & No observed spectral features                           \\
\enddata
\tablecomments{Objects identified as high redshift candidates are marked with an asterisk.}
\end{deluxetable}

\subsection{GOODS-North Candidate Spectra and Final Selection}
\label{sec:fcands}

Since all the objects presented in table \ref{tab} satisfy our primary
$i'-z'$ color selection criterion and do not have $z'-HK'$ colors
indicative of being lower redshift elliptical galaxies, our most
significant remaining source of contamination in a high redshift
sample is likely to be cool galactic stars. These constitute two of
the nine $i$-drop objects in the GOODS-S field (paper 1) and are often
excluded from colour selected samples by the simple expedient of
rejecting all unresolved sources (e.g. Bouwens et al.\ 2003). In this
section we consider the interpretation of the candidate spectra for
those objects for which such is available, the probable classification
of each candidate in table \ref{tab} and the construction of a robust
sample of high redshift galaxies.

Five objects in GOODS-N were targeted for spectroscopic
observations. Two yielded continuum detections, and a likely
identification as cold stars (candidates 1 and 2). One spectrum
(candidate 6) shows a single extended emission line feature at 8804\,\AA
, with the peak emission spatially offset from the continuum location by
$\sim 2\arcsec$ (and hence this spectral feature may be associated with
an adjacent foreground object). No signal was detected in the other two
spectra (candidates 5 and 10). A description of each spectrum follows.

\subsubsection{GOODS-N $i'$-Drop \#1}

The spectrum of GOODS-N $i'$-drop \#1 does not appear to show any
distinctive emission lines. The object continuum, however, is clearly
detected at long wavelengths.  Figure \ref{fig:c12} presents a box-car
smoothed spectrum of this object together with the spectrum of an
L-class star (2MJ0345432+254023, class L0 for comparison.

The object's spectrum and compactness R$_h$ =0\farcs05 (unresolved)
are both strongly suggestive that this object may be identified as an
M or early L class star and although the spectrum obtained has very
low signal-to-noise, there are suggestive commonalities with the L0
spectrum such as the absorption feature at $\sim$8400\AA.  In addition
its colors are marginally consistent with identification as a mid
M-class star, although the near-infrared colors are affected by a near
neighbor.  We provisionally identify this unresolved source as a
galactic star, and hence exclude it from our final high redshift
sample.

\subsubsection{GOODS-N $i'$-Drop \#2}
 
\begin{figure}
\plotone{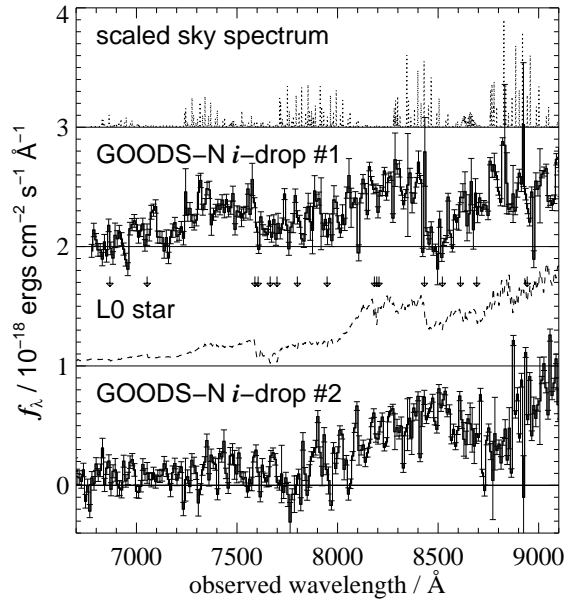}
\caption{GOODS-N $i'$-drops \#1 and \#2 --- Spectra are binned into
10\,\AA\ bins (33\,pixels) and are extracted over 1\arcsec\
(9\,pixels). The spectrum of candidate 2 is offset by $2\times
10^{-18}\,\mathrm{ergs\,cm}^{-2}\,\mathrm{s}^{-1}$ for clarity. The
spectrum of the L0 Dwarf star 2MJ0345432+254023 (2MASS, Reid et al.\
1999) is plotted for comparison (offest by $1\times
10^{-18}\,\mathrm{ergs\,cm}^{-2}\,\mathrm{s}^{-1}$), along with the sky
spectrum (both with arbitrary scaling). The primary spectral features of L
class stars are marked with arrows and correspond to the following lines
respectively: TiO, TiO, \mbox{K\,{\sc i}}, \mbox{K\,{\sc i}},
\mbox{Rb\,{\sc i}}, \mbox{Rb\,{\sc i}}, \mbox{Na\,{\sc i}},
\mbox{Na\,{\sc i}}, TiO, TiO, \mbox{CS\,{\sc i}} (Kirkpatrick et al.\
1999). }
\label{fig:c12}
\end{figure}

As in the case of GOODS-N $i'$-drop \#1, the spectrum of GOODS-N
$i'$-drop \#2 does not appear to show any distinctive emission lines
but the object continuum is clearly detected at long
wavelengths. Figure \ref{fig:c12} presents a smoothed spectrum of this
object.  Both the continuum colors of this object and its compactness
(R$_h$ =0\farcs05, unresolved) suggest that this object may be an L
dwarf star, although it lacks the absorption at $\sim$8400\AA\ often
seen in L class stars.  The low signal-to-noise spectrum of this
object shows a moderate cross correlation with those of L class stars
and a weaker correlation with the composite LBG spectrum of Shapley et
al.\ (2003), redshifted to $z\sim6$. We tentatively identify this
object as a cool galactic star and it is therefore excluded from our
final list of good high redshift candidates. As the classification of
this object is uncertain it should be noted that, as one of the
brighter objects selected by our $(i'-z')_{\rm AB}$ color criterion,
this object would have a significant effect on our calculations in
section \ref{sec:sfr} if it was shown to lie at high redshift, as is
discussed in that section.

\subsubsection{GOODS-N $i'$-Drop \#5}
 
Although a spectrum was obtained for GOODS-N $i'$-drop \#5, no
spectral features or continuum flux were observed.  However this
object is relatively faint ($z'_{\rm AB}=24.6$) and so we would be
unlikely to detect continuum emission. Our limit on the equivalent
width of an emission line is $W_{obs}<10$\,\AA\ (5\,$\sigma$,
300\,km\,s$^{-1}$ width), provided it lies away from a sky line.
GOODS-N $i'$-drop \#5 is relatively extended and so unlikely to be a
galactic star. Its near-infrared color is also inconsistent with
identification as an elliptical galaxy at $z\sim2$ although the colors
are not fully consistent with a high redshift identification either.
We include candidate 5 in our list of high redshift candidates
although we note the ambiguity in its near-infrared colors.

\subsubsection{GOODS-N $i'$-Drop \#6}

\begin{figure}

\resizebox{0.4325\textwidth}{!}{\includegraphics{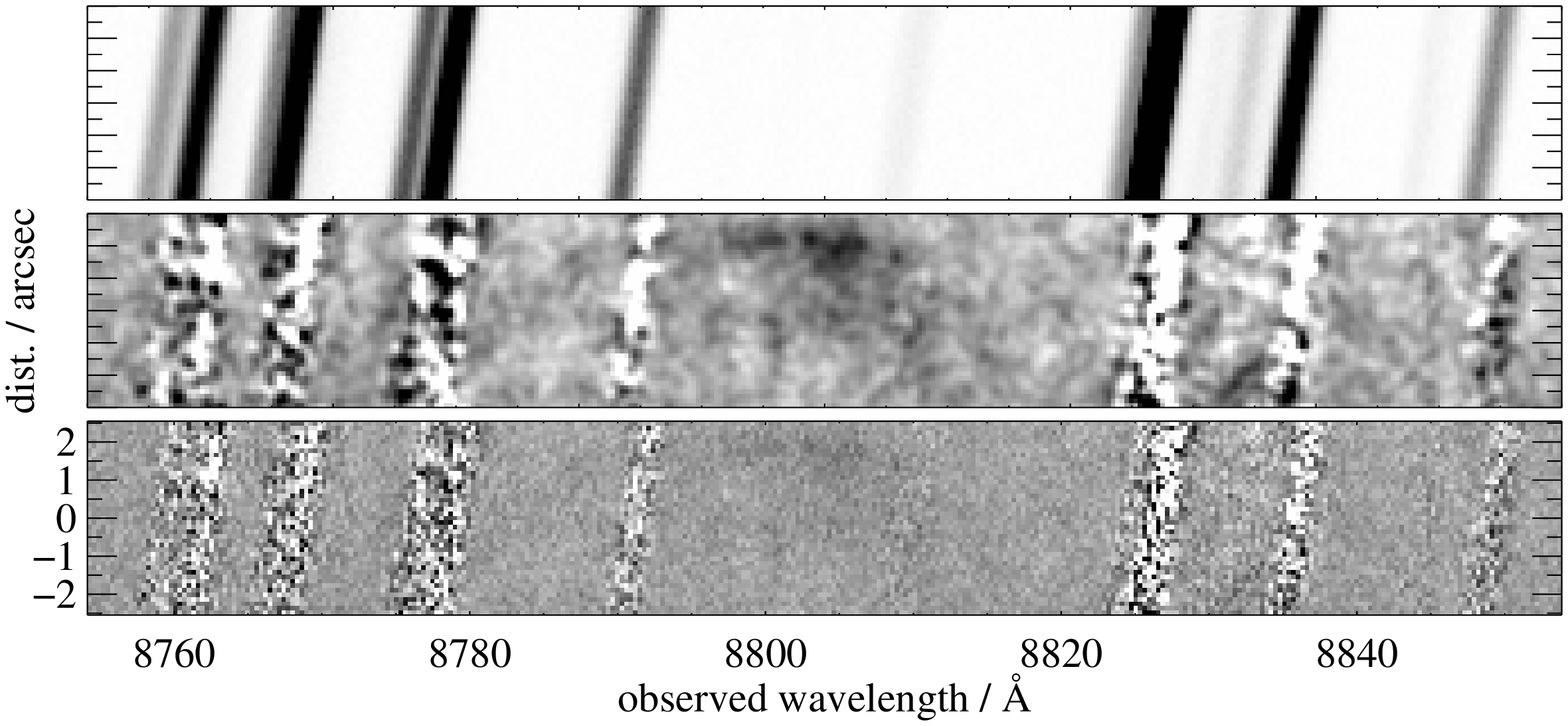}}
\resizebox{0.04\textwidth}{!}{\includegraphics*[20,0][48,140]{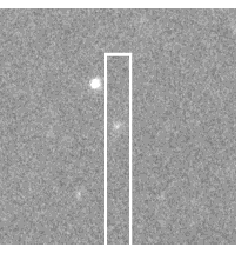}}

\epsscale{0.8}
\plotone{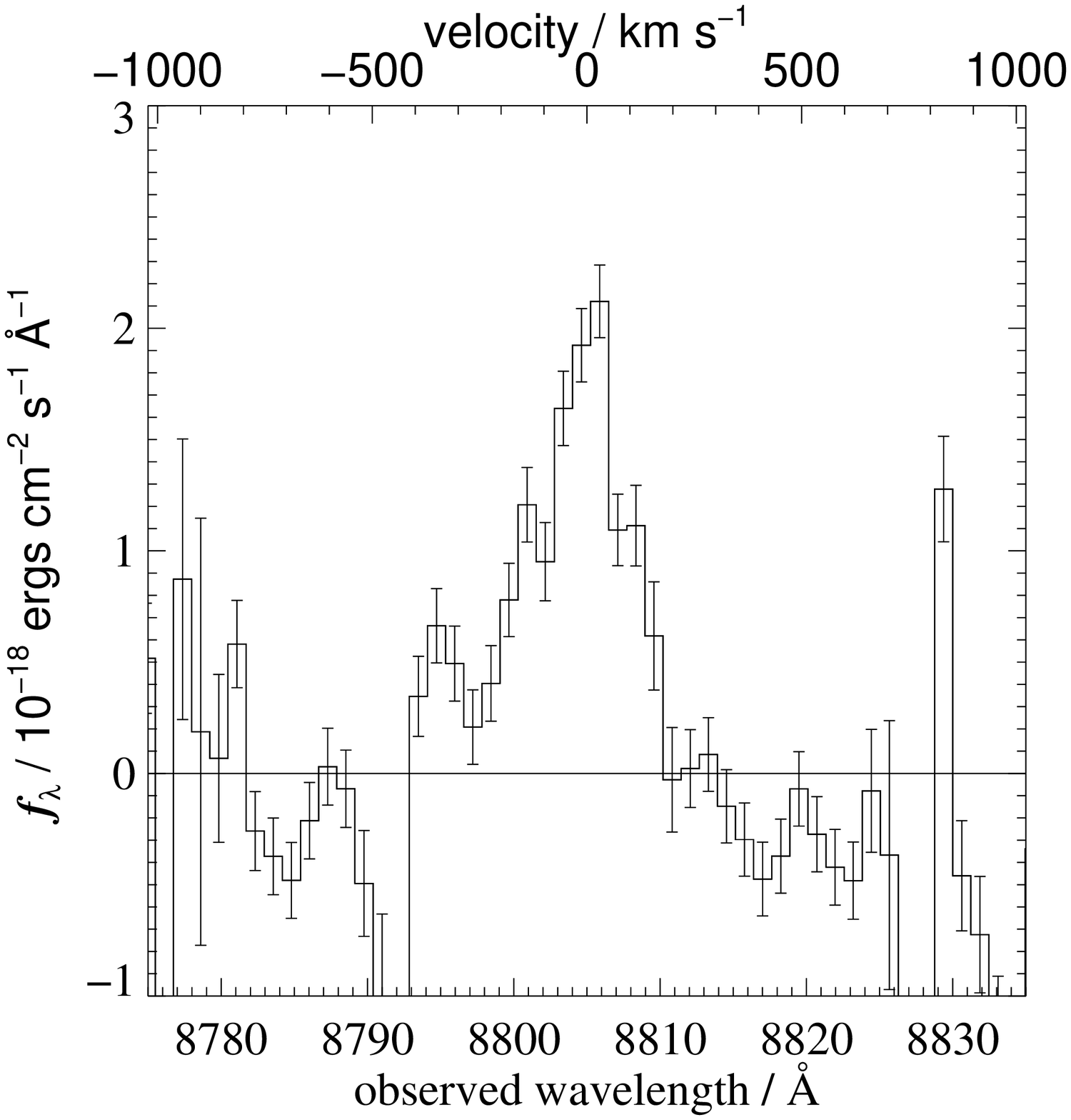}\\

\caption{(a) GOODS-N $i'$-drop \#6 2D spectrum --- Candidate Lyman-$\alpha$ emission
line at $z=6.24$ (lower panel),
sky spectrum (upper panel), 2D spectrum boxcar smoothed over 3 pixel
resolution elements (centre panel). The 2D spectra span a spatial range of $\pm
2.5\arcsec$ around the $i'$-drop location on slit. The emission line is
centered 1\farcs5 from the candidate object but extends to its
location. (Right) HST/ACS image showing the slit orientation. Box length is 10\arcsec. As can be seen the neighbouring object lies just off the slit but some flux from this neighbour would be expected to fall down the slit in ground-based seeing. (b) The 1D spectrum
extracted over 2\arcsec\ around the emission
line center, and binned into independent spectral resolution elements of
1.2\,\AA\ (4\,pixels). The features around 8790\,\AA\ and 8830\,\AA\ are
sky-subtraction residuals.}
\label{fig:c6}
\end{figure}

The 2D spectrum of GOODS-N $i'$-drop \#6 showed one significant emission line
 detected at $\approx 20\,\sigma$ at a central wavelength of
 $\lambda_{\rm obs}\sim8804\pm 1$\,\AA.  Both 2D and 1D extracted
 spectra for this line are shown in figure \ref{fig:c6}, centered on the
 spatial location of the $i'$-drop object and the central wavelength of
 the emission line. The emission line is spatially extended,
 stretching from the location of the $i'$-drop in the {\em ACS} imaging to
 peak in intensity  approximately $2\arcsec$ away .

The line flux is $(7\pm 2) \times
10^{-18}\,\mathrm{ergs\,cm}^{-2}\,\mathrm{s}^{-1}$ measured between the
zero power points at 8798.0\AA\ and 8813.5\AA\ in an extraction width of
$3\arcsec$ centered on the peak of the line emission. As figure
\ref{fig:c6} shows, the line profile is clearly resolved and shows
velocity structure, although it is not obviously a doublet.  The FWHM of
the emission line is $(10.0 \pm1.0)$\AA\ although the central peak of
emission is narrower with a FWHM of $(4.0\pm0.25)$\AA.

GOODS-N $i'$-drop \#6 is a relatively extended, faint galaxy with a brighter
neighbor at approximately $2\arcsec$ separation along the slit
direction.  The neighboring object is expected to lie just off the slit,
although given the seeing conditions of this observation
($\sim0\farcs8$) we may expect some light from this object to fall into
our slit near the location of the emission line. As a result we suggest
two possible interpretations of this observation.

One possible scenario is that the line emission is associated with
this lower-redshift (non-$i'$-drop) neighboring galaxy. The observed
line is isolated, with no other lines detected. The neighbor has a
photometric redshift of $z=2.605\pm0.2$, based on its colors in the
GOODS/{\em ACS} imaging and determined using the template fitting routine
`{\em HyperZ}' (Bolzonella et al.\ 2000). The observed line cannot be
easily identified with any emission line at $z=2.6$ ($\lambda_{\rm
rest}\sim2442$\AA ). It is inconsistent with identification as a
$z=1.36$ [O\,{\sc ii}] ($\lambda_{\rm rest}=3726, 3729$\AA ), and no
doublet structure is obvious, although the line could conceivably be
Mg\,{\sc ii} ($\lambda_{\rm rest}\sim2795$\AA ) at $z=2.15$.

The second scenario is that we are seeing spatially-extended
Lyman-$\alpha$ emission at $z=6.24$, as has been observed in a number of
star forming galaxies.  Observations of a lensed $z=4$ galaxy by Bunker
et al.\ (2000) identified Ly-$\alpha$ emission offset by 1-2\arcsec\
from the rest frame UV continuum region (although this corresponds to
only 0.7$\,h_{70}^{-1}\,$kpc at $z=4$ given the lensing amplification of
$\sim 10$ for this galaxy).  Similarly, Steidel and coworkers (2000)
identify two Lyman-$\alpha$ `blobs'
associated with but not centered on
continuum emission Lyman break galaxies at $z\sim3$.  These
Lyman-$\alpha$ regions extend over $>100\,h_{70}^{-1}\,$kpc and are
similar to extended Lyman-$\alpha$ emission seen in the vicinity of high
redshift radio-galaxies.

At $z\sim6$ a spatial offset of 2 \arcsec\ corresponds to a physical
separation of $11.4\,h^{-1}_{70}$\,kpc, larger than that observed in the
$z=4$ galaxy but much smaller than the extended Lyman-$\alpha$ observed
by Steidel et al.\ (2000). 
If we identify the emission line as Lyman-$\alpha$
then the emission wavelength is $\lambda_{\rm rest}\sim1216$\AA\ giving
a redshift for candidate 6 of $z=6.24$. We tentatively adopt this
identification and candidate 6 remains part of our high redshift galaxy
sample.

\subsubsection{GOODS-N $i'$-Drop \#10}
 
{GOODS-N $i'$-drop \#10 was selected with marginal $i'-z'$ colors,
slightly below our strict color cut criterion. The spectra of this
object shows no significant features or continuum flux in its
spectrum.  Both its compact (unresolved) $R_h$ and $(z'-HK')_{\rm AB}$
color suggests that this may be a low mass star. Given that this
object did not meet our strict criteria, the relative brightness and
its lack of strong high redshift emission features we exclude this
object from our final selection as a probable low mass star although,
again, we consider the effects of including this object in section
\ref{sec:sfr}.

\subsubsection{GOODS-N $i'$-Drops for Which No Spectra Are
Available} 

In common with GOODS-N $i'$-drops \#2 and \#10, GOODS-N $i'$-drop \#3 is
unresolved and its $z'-HK'$ color is completely consistent with
identification as an L-class dwarf star (Reid et al.\ 2001).  It is thus
excluded from the final robust high redshift sample although the effects of including all our stellar candidates as possible high redshift objects is discussed in section \ref{sec:sfr}. 

GOODS-N $i'$-drops \#4 and \#8, on the other hand, are relatively
extended objects and so are inconsistent with identification as galactic
stars. Their near-infrared colors are also inconsistent with elliptical
galaxies at $z\sim2$ suggesting that they may safely be identified as
high redshift candidates. These candidates remain in our selection.

GOODS-N $i'$-drop \#9 is relatively extended and can be identified with
Fern\'andez-Soto, Lanzetta and Yahil (1999) object 311 which was given a
photometric redshift of $z_{\rm phot}=5.64$ by those authors, consistent
with our result. This is the only high-redshift candidate object here to
appear in the Fern\'andez-Soto, Lanzetta and Yahil (1999) photometric
redshift catalogue for the HDFN. Since GOODS-N $i'$-drop \#9 is inside
the central Hubble Deep Field North region, it is observed in deep
NICMOS near-infrared imaging (Thompson et al.\ 1999), as shown in figure
\ref{fig:c9}, from which we have obtained this object's near infrared
magnitudes in the F110W$_{\rm AB} (J)$ and F160W$_{\rm AB} (H)$
bands. Its colors of $(z'-J)_{\rm AB}\sim0.53$ and $(J-H)_{\rm
AB}\sim0.50$ are consistent with a high redshift interpretation and
inconsistent with the colors of low redshift galaxies.  As a result we
include it in our list of final candidates.
\begin{figure}
\epsscale{0.32}
\plotone{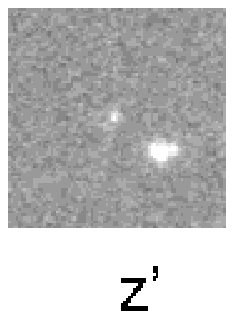}
\epsscale{0.64}
\plotone{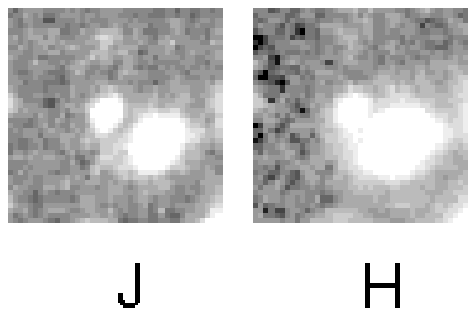}
\caption{GOODS-N $i'$-drop \#9 in the {\em ACS} $z'$-band image and in
the NICMOS F110W$_{\rm AB} (J)$ and F160W$_{\rm AB} (H)$ imaging of the
Hubble Deep Field North.}
\label{fig:c9}
\end{figure}

Since GOODS-N $i'$-drop \#11 was selected with marginal $(i'-z')_{\rm
AB}$ colors, not meeting the strict $(i'-z')_{\rm AB}<1.5$ color cut,
and is also unresolved it has also been rejected as a probable low mass
star despite the lack of a strong near-infrared color constraint. This
is the only object excluded from our final candidate selection primarily
on grounds of its unresolved nature. As such we note that its inclusion
in our final selection would have increased the observed star formation
rate of that sample by 18\% (see section \ref{sec:sfr}).

\subsection{Spectroscopy of $i'$-drops in GOODS-South Field}
\label{sec:CDFSspec}

\subsubsection{GOODS-S SBM03\#1} 

Our 5.5-hour spectrum of GOODS-S $i'$-drop \#1 in paper 1 (hereafter
SBM03\#1) reveals a single emission line at $8305.3\pm 0.9$\,\AA\ at the
same location on the 1\arcsec -wide longslit as the $i'$-drop
($\alpha_{2000}=03^h32^m40.0^s$,
$\delta_{2000}=-27^{\circ}48'15.0''$). The emission line falls between
two sky lines, and the 2D spectrum is shown in
figure~\ref{fig:SBM03_1_2D}.

The spectrum is very similar to that of GOODS-S SBM03\#3 reported in
Paper~2, where we argued that the most likely interpretation of the solo
emission line at 8245\,\AA\ was Lyman-$\alpha$ at $z=5.78$, given the
$i'$-drop selection and the fact that we would resolve the [OII] doublet
at lower redshift (and most other prominent emission lines would have
neighboring lines within our wavelength range). Applying the same logic
to our spectrum of GOODS-S SBM03\#1, it seems likely that the emission
line is Lyman-$\alpha$ at $z=5.83$ -- a view supported by the
characteristic asymmetry in the emission line spectrum, with a sharp
blue-wing cut-off\footnote{This object has been independently confirmed as lying at $z=5.8$ by Dickinson et al.\ (2003) and corresponds to their object SiD002.}. 
The redshift of SBM03\#1 is very close to the
redshift of SBM03\#3 (the line centers are separated by
2000\,km\,s$^{-1}$), and the spatial separation of 8\,arcmin corresponds
to a projected separation of only $2.9\,h_{70}^{-1}$\,Mpc. 

The flux in this emission line is comparable to that of SBM03\#3 at
$z=5.78$, and is within 30 per cent of $f=2\times
10^{-17}\,{\rm erg\,cm^{-2}\,s^{-1}}$, extracting over 17pixels
(2\,arcsec) and  measuring between the zero-power points
($8300-8318$\,\AA ). This is a lower limit due to potential slit losses,
although these are likely to be small, since like SBM03\#3, SBM03\#1 is
compact ($R_{\rm hl}=0.1\arcsec$, paper 1). After deconvolving with the
{\em ACS} $z'$-band point spread function, the half-light radius
corresponds to a physical size of only $0.5\,h^{-1}_{70}$\,kpc.

In continuum, SBM03\#1 is half the brightness of SBM03\# (paper 2) from
the $z'$-band magnitude. The equivalent width is $W_{\rm rest}^{\rm
Ly\alpha}=30\pm 10$\,\AA , at the upper end of the distribution observed
in $z=3-4$ Lyman break galaxies (Steidel et al.\ 1999). The star
formation rate from the rest-frame UV continuum is
$17\,h^{-2}_{70}\,M_{\odot}\,{\rm yr}^{-1}$ (paper 1).

\begin{figure*}
\epsscale{0.6}
\plotone{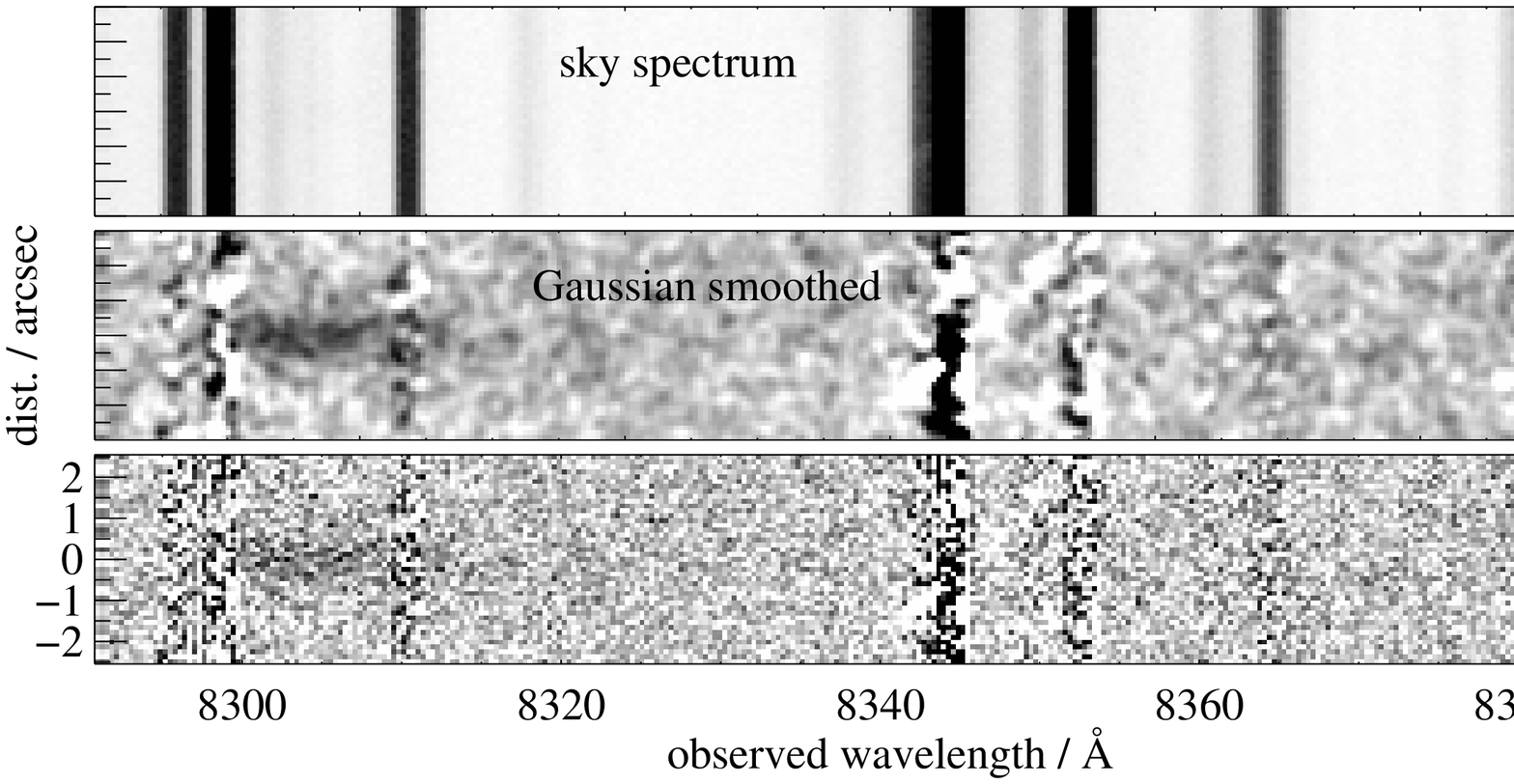}
\hspace{1cm}
\epsscale{0.35}
\plotone{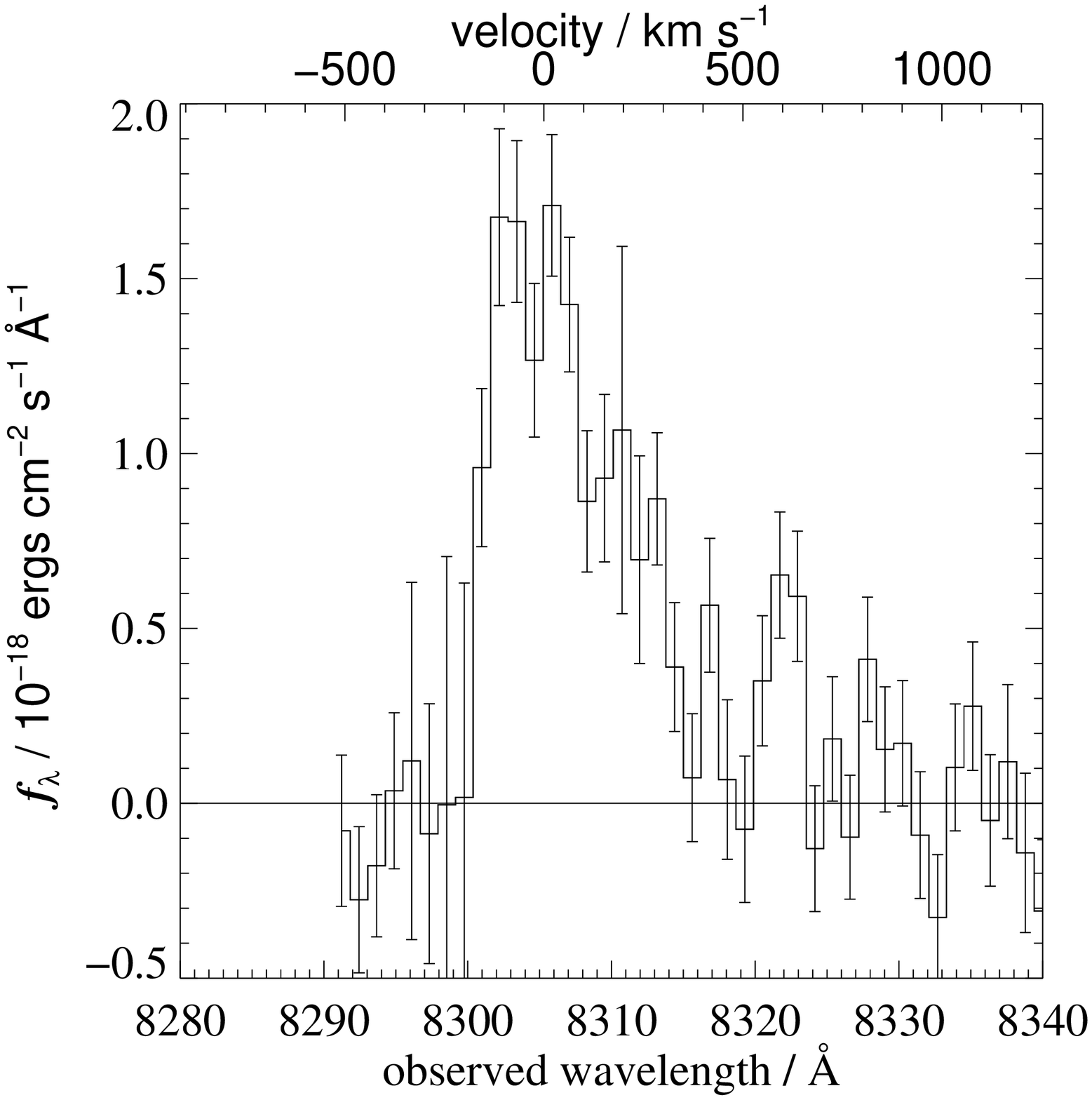}
\label{fig:SBM03_1_2D}
\label{fig:SBM03_1_1D}
\caption{(a) The lower panel shows the two-dimensional spectrum of
GOODS-S SBM03\#1 (candidate 1 in paper~1) around Lyman-$\alpha$
(100\,\AA\ by 5\,arcsec), the middle panel shows this smoothed with a
Gaussian of $\sigma=1$\,pixel, and the upper panel shows the sky
spectrum for this range. The line is centered at 8305\,\AA . (b) The
one-dimensional spectrum of SBM03\#1 around Lyman-$\alpha$, extracted
over a 9pixel (1\,aecsec) width. The data have been binned into
independent spectral resolution elements of 1.3\,\AA\ (4\,pixels).}

\end{figure*}

\subsubsection{Further Candidates in GOODS-S}

As described in Paper 2, DEIMOS spectroscopy of three further
candidates (SBM03\#2,5,7) was obtained with the Keck II telescope.
Inspection of these spectra reveals no obvious Lyman $\alpha$ emission
or continuum emission within the wavelength range sampled. At present
the nature of these sources is unfortunately unclear.

\section{High Redshift Galaxies - Derived Parameters and Interpretation}
\label{sec:discussion}

\subsection{Survey Volume and Completeness} 
\label{sec:volume}

Galaxies in the redshift range $5.6<z<7.0$ are selected by our
color-cut provided they are sufficiently luminous. However, as we
discuss in paper 1, our sensitivity to star formation is not uniform
over our survey volume: at higher redshifts, we are only sensitive to
more luminous galaxies at our apparent magnitude limit of
$z'_{AB}=25.6$, exacerbated by the increasing fraction of the
$z'$-band extincted by the Lyman-$\alpha$ forest, and there is a
$k$-correction as we sample progressively shorter rest-frame
wavelengths. As a result of this luminosity bias, the average redshift
of objects satisfying our selection criteria is towards the lower end
of the redshift range above and we calculate this luminosity-weighted
central redshift as $z=5.8$ under the assumption of no evolution in
the luminosity function from that at $z\sim3$.  In paper 1 we followed
the methodology of Steidel et al.\ (1999) to quantify the
incompleteness due to luminosity bias with redshift by calculating the
effective volume of the survey. If there is no change in
$L^*_{LBG}$(1500\,\AA ) or $\alpha$ from $z=3$ to $z\approx 6$
(i.e. no evolution in the galaxy luminosity function from that of
Steidel et al. 1999 at $z=3$), we showed that for our {\em ACS}
$z'$-band limit for the GOODS data, the effective volume is 40 per
cent that of the total volume in the interval $5.6<z<7.0$. For our
200\,arcmin$^{2}$ survey area (including areas only surveyed in half
of the epochs and after excluding the unreliable regions close to the
CCD chip gaps) this corresponds to an {\em effective} comoving volume
of $V_{\rm eff}=2.5\times 10^{5}\,h^{-3}_{70}\,{\rm Mpc}^{3}$. This
assumes a spectrum flat in $f_{\nu}$ at wavelengths longward of
Lyman-$\alpha$ (i.e., $f_{\lambda}\propto \lambda^{-\beta}$ where
$\beta=2.0$). For a redder spectrum with $\beta=1.1$ (the mean
reddening of the $z\approx 3$ Lyman break galaxies, Meurer et al.\
1997) the effective volume is 36 per cent of the total ($V_{\rm
eff}=2.3\times 10^{5}\,h^{-3}_{70}\,{\rm Mpc}^{3}$ comoving) -- which
would result in an 8\% rise in the measured star formation density
(section \ref{sec:sfr}) compared with $\beta=2.0$.

The above discussion, however does not take into account the effects of
 survey incompleteness due to photometric scatter, extreme intrinsic
 color variation or cosmological surface brightness dimming.

Our selection procedure is robust to the effects of photometric
scatter. By utilizing four epochs of data and selecting objects in
each which satisfy our criteria, only those objects lying more than
1$\sigma$ from their true magnitudes in four independent measurements
will be entirely missed (fewer than 1\% of the population assuming
Gaussian errors on the magnitudes).  It is possible, however, that the
effects of photometric errors in one or more epochs may sufficiently
affect the combined flux of the observations to take an object below
our $(i'-z')_{\rm AB}$ selection limit.  In practise only two objects
in our epoch-by-epoch selection sub-catalogue showed $(i'-z')_{\rm
AB}$ colors that consistantly exceeded $(i'-z')_{\rm AB}>1.3$ and
fluctuated around our color cut and these were included in table
\ref{tab} as candidates 10 and 11.  Both these objects were excluded
from the final high redshift candidate selection on the basis of their
stellar colors and full-width half-maxima.  Since relaxing our colour
cut criterion to $(i'-z')_{\rm AB}>1.3$ does not significantly add to
the number of candidate objects, while at the same time increasing the
likely fraction of low redshift contaminants (see figure
\ref{plot:tracks}), we estimate that the fraction of objects falling
out of our selection due to photometric errors is small.

The effect of extreme scatter in the distribution of intrinsic object
colors on the completeness of our selection is difficult to quantify
given the unknown nature of the $z\sim6$ population.  Our color
selection criterion illustrated in figure~\ref{plot:tracks} is based
upon the colors predicted by the evolution of empirical galaxy
templates. This procedure has been shown to work well in the selection
of Lyman break galaxies at $z\sim3-4$ but does not account for
atypical galaxies or for any significant deviation in the spectral
shape of very high redshift objects from the templates.

As described in section \ref{sec:phot}, simulations were made to
assess the completeness of our $i'-z'$ colour selection criterion on a
galaxy population with the characteristics of the $z\sim3$ Lyman break
galaxies (Adelberger \& Steidel 2000). These found that $>99$\% of
$z\ge5.7$ galaxies brighter than our luminosity limit in such a
population would be identified using this technique, with a somewhat
lower completeness in the $5.6<z<5.7$ redshift bin where the
completeness depends sensitively on the filter transmission and the
model of IGM absorption.  Relaxing the colour cut criterion to
$(i'-z')_{\rm AB}>1.3$ would increase our completeness in the
$5.6<z<5.7$ redshift bin by $\approx10\%$ without affecting
completeness in the higher redshift bins but, as noted above, would
also slightly increase the levels of contamination from $z\sim2$
elliptical galaxies.

The characteristics of the $z\sim6$ population are currently unknown
with neither the luminosity function or the distribution of spectral
slopes accessible to the available data and it is possible that there
is significant evolution in these properties in the interval $3<z<6$.
However, the $i'-z'$ colour of galaxies at these high redshifts is
driven primarily by the break caused by absorption in the intervening
intergalactic medium.  As a result the $i'-z'$ colour is effectively
independent of spectral slope and the colours of a wide variety of
intrinsic spectral energy distributions converge for $z>5$.

The single object confirmed to lie within our redshift range in the
GOODS-N field (HDF 4-473 at $z=5.60$, Weymann et al.\ 1998) is
slightly too faint to appear in our object selection ($z'_{\rm
AB}=26.66$, Bouwens et al.\ 2003a).  More significantly, this galaxy
is also too blue to appear in our selection ($(i'-z')_{\rm AB}=1.2$,
Bouwens et al.\ 2003a), having been observed as a $V$ band drop-out,
and so raises the possibility that we are excluding real objects by
applying our color-cut.  We note that HDF 4-473 is at the very lower
end of the redshift range to which we are sensitive, and hence in a
bin where our completeness is relatively poor.  Nonetheless this
object highlights the caution which must be applied to the
interpretation of any sample selected purely on photometric criteria.

\subsection{Physical Sizes of the $i'$-drops and Surface Brightness
Selection Effects} 
\label{sec:sizes}

Our final catalog selection includes objects with observed sizes in the
range R$_h\sim0\farcs08-0\farcs20$, similar to the observed range of
sizes in the 6 $z\sim6$ candidates we presented in paper~1. This range
corresponds to projected physical sizes of $0.3-1.0\,h^{-1}_{70}\,$kpc
at $z\sim5.8$, deconvolved with the instrument PSF. The size range of
our total sample of 12 $z\sim6$ objects (paper 1 + this work) is
comparable to, or smaller than, those of $z>5$ galaxies reported by
Bremer et al.\ (2003). Bremer and coworkers identified a sample of 44
sources with half-light radii of R$_h\sim0\farcs1-0\farcs3$ and
suggested that, when taken in conjunction with observations at lower
redshifts, their results implied a modest decrease in galaxy scale
lengths with redshift. Our results appear to support this conclusion,
although it should be noted that at $z'_{\rm AB}=25.6$, objects more
extended than R$_h\sim0\farcs25$ ($1.5\,h^{-1}_{70}\,$kpc) fall out of
our selection due to low surface brightness (as shown in figure
\ref{fig:zfwhm}), thus placing a significant observational selection
effect on the observed distribution of sizes.  

Simulations were performed using the {\sc IRAF.Artdata} package,
placing artificial galaxies in the GOODS images in order to test the
recoverability of such objects given our SExtractor detection
parameters.  We model the input population as having exponential disk
surface brightness profiles with half light radii distributed
uniformly between the instrumental point spread function (0\farcs05)
and 0\farcs5.  We find that we reliably recover $>$95\% of the input
galaxies for half-light radii R$_h<0\farcs25$ and $>$98\% of galaxies
with R$_h\sim0\farcs20$ with the remaining 2\% being lost because of
crowding.

Nonetheless, at brighter magnitudes our objects do not occupy the full
range of half-light radii allowed by the observational constraint: at
$z'_{\rm AB}=24.7$ (the magnitude of our brightest high redshift
candidate) we are sensitive to objects as large as R$_h\sim0\farcs34$
(see figure \ref{fig:zfwhm}).  This suggests that most objects at high
redshift are small enough to fall within our surface brightness
selection limit.

Star forming regions in the local universe are often very compact in
the UV continuum and it is likely that this is also true at high
redshift.  Lowenthal et al.\ (1997) report that Lyman break galaxies
at $z\sim3$ are compact and, if projected to $z\sim3$, this population
would range in half-light radius from 0\farcs08-0\farcs35 with a
median size of R$_h=0\farcs18$. Although just over half of this range
of half-light radii would be detectable at our magnitude limit, the
full range of half-light radii are accessible to this sample at
brighter magnitudes and failure to detect such objects may indicate a
reduction in the physical scale of galaxies with increasing
redshift. At $z>5$ Bremer and coworkers (2003) identified only 2
candidate objects on scales significantly larger than
R$_h\sim0\farcs3$ ($1.8\,h^{-2}_{70}$\, kpc at $z\sim5.2$) in a sample
of 44 galaxies and considered these objects most likely to be $z\sim1$
red galaxies. Iwata et al.\ (2003) made a similar selection of Lyman
break galaxies at $z>5$ and found 40 per cent of their candidate
objects were unresolved although they were limited by seeing
conditions in their ground based observations.  As Bremer et al.\
discuss, they expect a contamination from lower redshift galaxies of
$\sim20\%$ in their sample and also expect such objects to be resolved
on scales of $>0\farcs3$.

While this constraint on the physical size of observed objects is not
explicitly accounted for in our effective volume calculation, the
change in the angular diameter of objects of the same physical size
over our redshift range is only 10 per cent.  Given the evidence for a
slight reduction in intrinsic object size with redshift (Bremer et
al.\ 2003, Roche et al.\ 1998), and the failure of our candidates to
occupy the full size range allowed by our observations even at fainter
magnitudes, we expect the fraction of objects lost due to surface
brightness effects to be small ($<2$\%) and the effect on our effective survey
volume to be negligible. 

\section{The Star Formation Rate at $z\sim 6$} 
\label{sec:sfr}

\subsection{The Global Star Formation History in GOODS-N} 
Since the primary sources of rest-frame ultraviolet photons within a
galaxy are short-lived massive stars, UV flux is a tracer of star
formation and its rest-frame UV flux density can be used to quantify the
total star formation rate (SFR) within a galaxy. This rest-frame
ultraviolet continuum is measured at $z\sim6$ by the $z'$ photometric
band.

The relation between the flux density in the rest-UV around $\approx
1500$\,\AA\ and the star formation rate (${\rm SFR}$ in $M_{\odot}\,{\rm
yr}^{-1}$) is usually assumed to be given by $L_{\rm UV}=8\times 10^{27}
{\rm SFR}\,{\rm ergs\,s^{-1}\,Hz^{-1}}$ (Madau, Pozzetti \& Dickinson
1998) for a Salpeter (1955) stellar initial mass function (IMF) with
$0.1\,M_{\odot}<M^{*}<125\,M_{\odot}$.

Our limiting star formation rate as a function of redshift was
considered in detail in paper 1, section 4 and illustrated in figure 6
in that paper.  Accounting for filter transmission, the effects of the
intervening Lyman-$\alpha$ forest and for small $k$-corrections to
$\lambda_{\rm rest}=1500$\,\AA\ from the observed rest-wavelengths our
limiting magnitude ($z'_{\rm AB}=25.6$) we should detect unobscured star
formation rates as low as $15\,[16.5]\,h^{-2}_{70}\,M_{\odot}\,{\rm
yr}^{-1}$ at $5.6<z<5.8$ and $21\,[25]\,h^{-2}_{70}\,M_{\odot}\,{\rm
yr}^{-1}$ at $z=6.1$ for spectral slope $\beta=-2.0\,[-1.1]$,
appropriate for an unobscured starburst and a redder slope appropriate
for mean reddening of the $z\approx 3$ Lyman break galaxies (Meurer et
al.\ 1997) respectively.

Steidel et al.\ (1999) used a spectroscopic sample of Lyman break galaxies
(LBGs) to constrain the luminosity function of these objects around
$\lambda_{\rm rest}=1500$\,\AA\ obtaining a characteristic magnitude at
the knee of the luminosity function of $m^{*}_{R}=24.48$ at $\langle
z\rangle=3.04$, with a faint end slope $\alpha=-1.6$ and normalization
$\phi^{*}\approx 0.005\,h_{70}^3\,{\rm Mpc}^{-3}$. A second sample of
galaxies at $\langle z\rangle =4.13$ was consistant with no evolution in this
luminosity function.  If we further assume that the luminosity function
does not evolve between $z\sim 3$ and $z\sim 6$ then our catalogue limit
at $z'_{\rm AB}<25.6$ would include galaxies down to $L^*_{\rm LBG}$
(corresponding to a characteristic star formation rate of ${\rm
SFR}^{*}_{\rm UV}=15\,h^{-2}_{70}\,M_{\odot}\,{\rm yr}^{-1}$) at $z\sim
6$ (although see section \ref{sec:lum} for a discussion of this
assumption).

In the ninth column of table \ref{tab} we use the ultraviolet flux ---
star formation rate relation assumed above to estimate the inferred
star formation rate for each of our $i'$-drop candidates assuming our
luminosity weighted central redshift ($z=5.8$) for each object.  In
section \ref{sec:fcands}, however, we commented on the nature of each
of the $i'$-drop objects and selected a subsample of six candidate
high-redshift star-forming galaxies with $z>5.6$. These have an
integrated star formation rate of $(135\pm
55)\,h_{70}^{-2}\,M_{\odot}\,{\rm yr}^{-1}$ (where the error is based
purely on Poisson statistics).  

If the luminosity function of Lyman break galaxies is assumed not to
evolve between $z=3$ and $z\sim6$ this gives a comoving
volume-averaged star formation density between $z = 5.6$ and $z=6.1$
of $\rho_{\rm SFR}=(5.4\pm 2.2) \times
10^{-4}\,h_{70}\,M_{\odot}\,{\rm yr}^{-1}\,{\rm Mpc}^{-3}$ for objects
with star formation rates $>15\,h^{-2}_{70}\,M_{\odot}\,{\rm
yr}^{-1}$. The inclusion of all objects in table 1 (the six objects in
our robust high redshift sample and the five probable stars) would
give a value some $4\times$ higher than this, with half of the
integrated star formation rate contributed by the unresolved low mass
star candidates \#1 and \#2 (see section \ref{sec:fcands}). While it
is true that none of our five unresolved candidates can be
conclusively identified as galactic stars, we are confident that this
figure would seriously over-estimate the true value given the
demonstrated stellar contamination of similar samples (e.g. paper 1).

Star formation densities calculated for both our high redshift galaxy
selection (filled square) and for all $i$-drops in table \ref{tab}
(open square) are compared to global star-formation rates found in
paper 1 and at lower redshifts in figure \ref{fig:madau}, adapted from
Steidel et al.\ (1999) and recalculated for our $\Lambda$ cosmology
and higher limiting star formation rate.  If our high redshift
candidate selection is adopted, then for Lyman break galaxies with
star formation rates $>15\,h^{-2}_{70}\,M_{\odot}\,{\rm yr}^{-1}$
(L$^{\ast}_{z=3}$) it appears that the observed comoving star
formation density was $\sim9$ times {\em less} at $z\sim 6$ than at
$z=3$ (based on the bright end of the UV luminosity function), a
conclusion reinforced by the close agreement between our results for
the GOODS-N and GOODS-S data.  If all $i$-drops in table \ref{tab} are
considered the fall in star formation density is less marked (a factor
of 2), however given the significant contribution the five unresolved
objects make to the star formation rate and the evidence against their
identification as high redshift objects, we are reluctant to bias our
results unduly high by their inclusion in the star formation density
estimate presented here.

The close proximity, both in velocity and in projected 
space, of the two confirmed $z\sim 5.8$ galaxies is suggestive of the 
presence of a group of galaxies at this redshift in the GOODS-S field.
 Nonetheless, the similar results obtained in both fields
suggests that our survey fields are sufficiently large that cosmic 
variance, a problem that has always plagued small deep-field surveys, 
does not contribute significantly towards the uncertainties associated 
with the determination. 

The observation that there is evolution in the luminosity function of
Lyman break galaxies in the redshift interval $3<z<6$ (as observed by
the bright end, SFR$>15\,h^{-2}_{70}\,M_{\odot}\,{\rm yr}^{-1}$) is
supported by simulation.  If an ensemble of galaxies with properties
identical to the $z\sim3$ Lyman break population in distribution of
spectral slope and luminosity function (Adelberger \& Steidel 2000) is
projected to lie at $z>4$ then we would expect to detect $49\pm10$
such galaxies at $z>5.6$ to our magnitude limit,
where the error bars incorporate contributions of similar magnitude
from poisson sampling noise and from uncertainties in the $z\sim3$
luminosity function. Hence the observed fall in number density of
bright galaxies is strongly suggestive of evolution in either the
normalisation ($\phi^*$, and hence star formation rate) or in the
shape ($L^*$ or $\alpha$) of the Lyman break luminosity function in
this interval (see section \ref{sec:lum} for further discussion).

Previous authors have integrated the star formation density over the
luminosity function for their objects down to star formation rates of
0.1\,$L^{*}$ (SFR$=1.5\,h^{-2}_{70}\,M_{\odot}\,{\rm yr}^{-1}$) even
if these SFRs have not been observed. We prefer to plot what we
actually observe (SFR $>15\,h^{-2}_{70}\,M_{\odot}\,{\rm yr}^{-1}$)
and not to extrapolate below the limit of our
observations. Unfortunately, calculation of the effective survey
volume requires an implicit assumption of the shape of the very
uncertain $z\sim6$ Lyman break galaxy luminosity function, as
discussed in section \ref{sec:volume}.  If the Steidel et al.\ (1999)
luminosity function for $z\sim3$ LBGs (used to calculate this volume)
is assumed to hold at $z\sim6$ and is integrated down to
$0.1\,L^*_{z=3}$ then the global SFR is increased by a factor of
$\sim5$ for $\alpha=-1.6$. See section \ref{sec:lum} for a discussion
of the effects of evolution in the luminosity function of these
objects.

\begin{figure}
\plotone{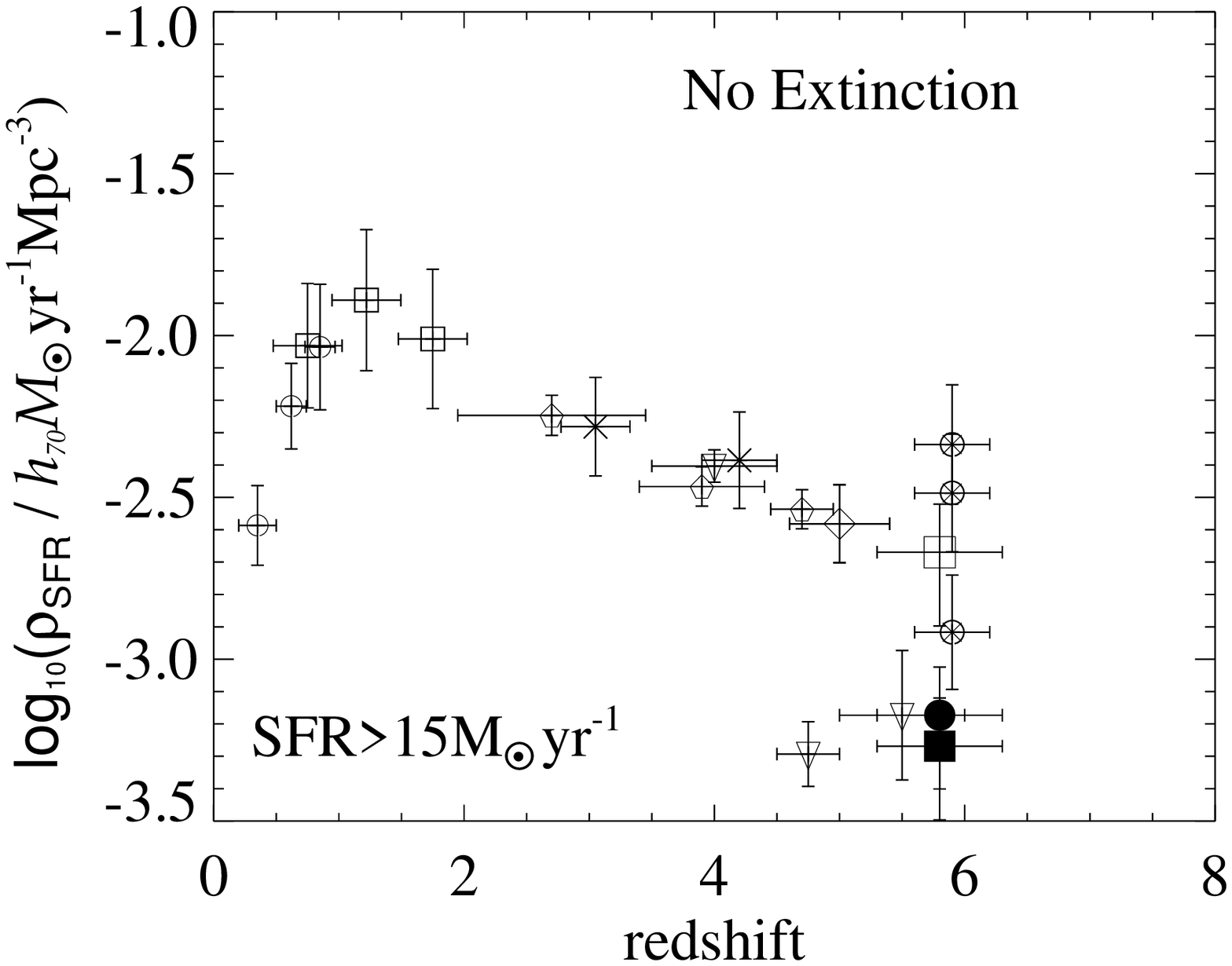}
\plotone{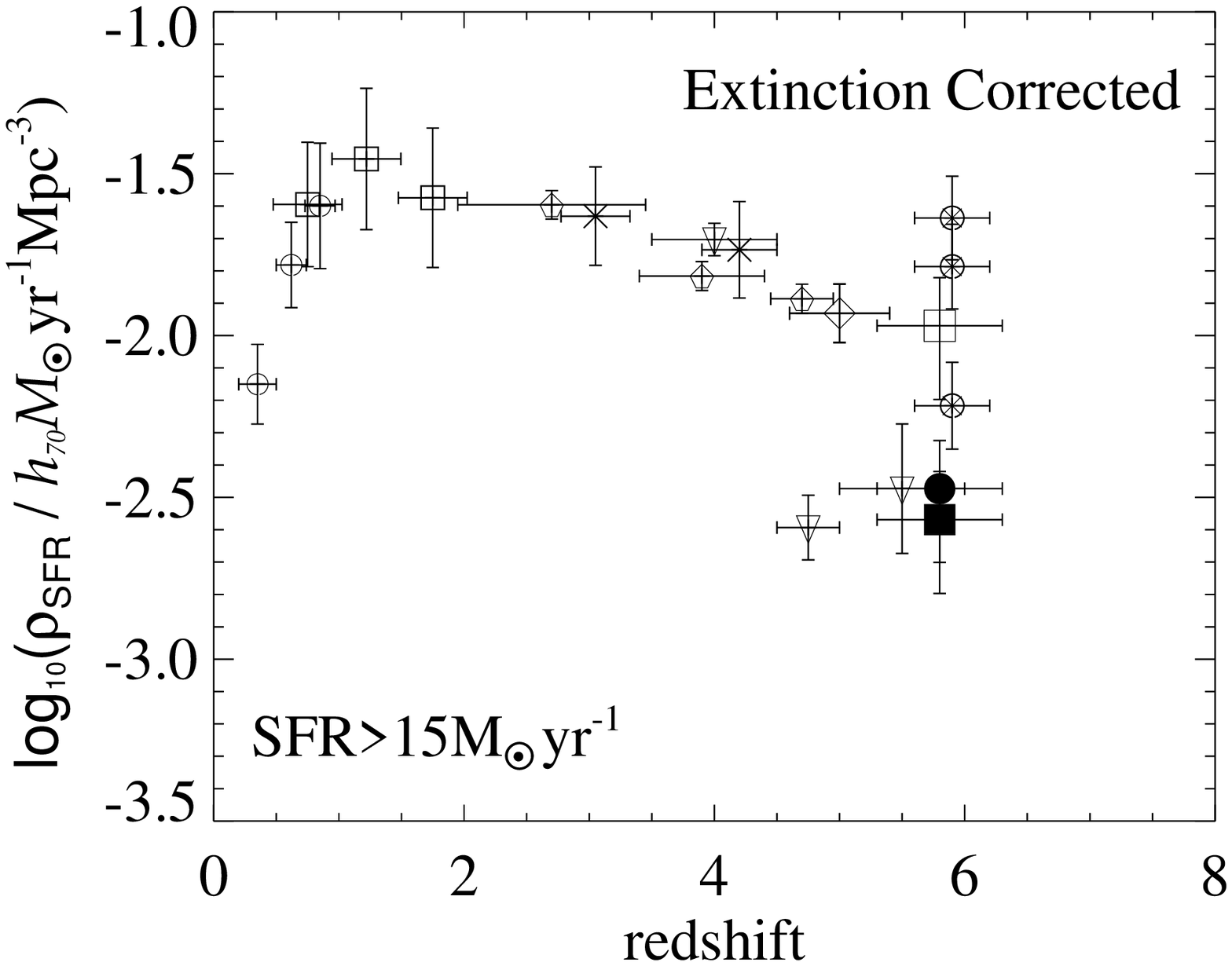}
\caption{The `Madau-Lilly' plot illustrating the evolution of the
comoving volume-averaged star formation rate.  Our work is plotted as
solid symbols: the result from paper 1 (GOODS-S) is shown as a filled
circle and from this work (GOODS-N) as a filled square (note that this
point includes only objects in our final high redshift selection). The
large open square at $z=5.8$ indicates the integrated star formation
rate in the GOODS-N if the five unresolved galactic star candidates
are assumed to be unresolved galaxies at $z\sim6$. Other determinations
have been recalculated for our cosmology and higher star formation rate limit
of $15\,h^{-2}_{70}\,M_{\odot}\,{\rm yr}^{-1}$ (instead of $1.5\,M_{\odot}\,{\rm yr}^{-1}$ for Steidel et al.\ 1999), assuming
a slope of $\alpha=-1.6$ for $z>2$ and $\alpha=-1.3$ for $z<2$. Data
from the CFRS survey of Lilly et al.\ (1996) are shown as open
circles; data from Connolly et al.\ (1997) are squares; and the Lyman
break galaxy work of Steidel et al.\ (1999) is plotted as crosses, of
Bouwens et al.\ (2003b) by pentagons, of Fontana et al.\ (2002) as
inverted triangles and that by Iwata et al.\ (2003) as an open
diamond. The three estimates of Bouwens et al.\ (2003a) are shown by
crossed circles and indicate three different completeness corrections
for one sample of objects (see discussion in
Section~\ref{sec:previous_studies}).  The lower panel indicates the
star formation density corrected for extinction using the Calzetti
(1997,2000) reddening law and the typical $E(B-V)=0.15$ value found by
Steidel et al.\ for their $z\sim3$ sample and applied by them at
$z\sim4$ and by Iwata et al.\ at $z\sim5$. Note that this assumption
of no evolution in the dust characteristics of Lyman break galaxies
between $z\sim3$ and $z\sim6$ is made primarily for comparison with
the work of previous authors.}
\label{fig:madau}

\end{figure}

\subsection{Extinction Corrections}
\label{subsec:dust}

The rest-frame UV is, of course, very susceptible to extinction by dust,
and such large and uncertain corrections have hampered measurements of
the evolution of the global star formation rate (e.g., Madau et al.\
1996).  Although extinction corrections increase the UV flux, and hence
the implied star formation rate, the method used here to select galaxies
(the Lyman break technique) has been consistently applied at all
redshifts above $z\approx 2$. Thus equivalent populations should be
observed at all redshifts and the resulting dust correction alters the
normalization of all the data points in the same way. Hence the fall in
the volume-averaged star formation rate between $z\approx 6$ and
$z\approx 3$ derived from our $i'$-drops does not depend on dust
extinction, unless the dust content is itself evolving lockstep with
redshift in the observed population (i.e., galaxies would have to be
more heavily obscured at early epochs to explain our results if the
global star formation rate remains roughly constant).

Vijh et al.\ (2003) made a study of the dust characteristics of
Lyman-break galaxies using the P\'EGASE synthetic galaxy templates.  They
report an ultraviolet flux attenuation factor due to dust of $5.7-18$ at
$2<z<4$ (c.f., Steidel et al.\ 1999, who derive a UV attenuation factor
of $4.7$ using the reddening formalism of Calzetti et al.\ 1994, 2000).
Vijh et al.\ find no evolution in dust characteristics of Lyman break
galaxies in this redshift range. Hence our assumption of no evolution in
the dust properties of Lyman break galaxies from $z\approx 3$ to
$z\approx 6$ (when the Universe was only $\sim$0.55 Gyrs old) may not be
unreasonable. The lower panel of figure~\ref{fig:madau} shows the
extinction-corrected volume-averaged star formation history.

\subsection{The Evolution of Lyman-$\alpha$ Emission?}
\label{subsec:lyman-alpha}

For the total sample (GOOD-S + GOODS-N), we confirm very few $i'$-drops
as being high-redshift galaxies in Lyman-$\alpha$ emission (2 definite
and one possible out of a total of 10 $i'$-drop objects observed
spectroscopically). However, we note that of the 10 $i'$-drops targetted
with Keck spectroscopy, two have stellar spectra (GOODS-N \#1 \& \#2)
and a further two have near-infrared colours of low-mass stars (GOODS-N
\#10; GOODS-S \#5), all four being unresolved point sources in the
HST/ACS images. So, excluding the probable stars, we have strong
Lyman-$\alpha$ emission in two out of six galaxies, one third of our
spectroscopic sample (SBM03\#1 $W_{\rm rest}^{\rm Ly\alpha}=30$\,\AA ,
this paper ; SBM03\#3 $W_{\rm rest}^{\rm Ly\alpha}=20$\,\AA ,
paper~2). This is in fact consistent with the fraction of strong
Lyman-$\alpha$ emitters at $z\sim 3$: Shapley et al.\ (2003) report that
25 per~cent of Lyman-break galaxies have Lyman-$\alpha$ in emission with
rest frame equivalent widths $>20$\,\AA .

Now we consider whether we would be sensitive to emission lines of lower
equivalent width in our spectroscopy. Our $5\,\sigma$ sensitivity to
line emission is $3\times 10^{-18}\,{\rm ergs\,cm^{-2}\,s^{-1}}$ (for a
1-arcsec extraction and a 300\,km\,s$^{-1}$ line-width). Therefore, at
the magnitude cut-off of our sample ($z'_{AB}=26.5$), this corresponds
to $W_{\rm rest}^{\rm Ly\alpha}>5$\,\AA\ at $z\approx 6$. The fact that
we do not observe any other Lyman-$\alpha$ emission lines, despite our
sensitive equivalent width limit, is marginally at odds with the global
properties of the Lyman-break sample at $z\sim 3$ reported in Shapley et
al.\ (2003): 20 per~cent of the Shapley sample have Lyman-$\alpha$ in
emission with  $W_{\rm rest}^{\rm Ly\alpha}<20$\,\AA\ -- so we might
expect to have seen another emission line in our sample at lower
equivalent width than the two emission lines found.

There is some evidence for a decline in the fraction of Lyman-$\alpha$
emitters at $z\sim 6$ compared with $z\sim 3$, although the statistics
are marginal. If this proves to be the case, it could imply that the
contamination of our total sample by low-$z$ galaxies is
significant, or perhaps that Lyman-$\alpha$ does not routinely escape
from $z\sim 6$ galaxies.  Lyman-$\alpha$ is seen in emission in half of
star forming galaxies at $z\sim 3$ (see Steidel et al.\ 1999). If, at
$z\sim 6$, we are at the end of reionization then neutral gas in the
Universe may resonantly scatter this emission (the Gunn-Peterson effect)
and so lead to a fall in the observed fraction of Lyman-$\alpha$
emitters.  This effect may have been detected by Becker et al.\ (2001)
at $z=6.3$ and could indicate either that reionization ended later than
estimated by the WMAP results (Kogut et al. 2003) or that pockets of
neutral gas remained well after the end of the bulk of reionization.

Maier et al.\ (2003) describe the Calar Alto Deep Imaging Survey (CADIS)
search for high redshift Lyman-$\alpha$ emitters.  They also find that
bright Lyman-$\alpha$ emitters are rarer at $z\ge5$ than suggested by a
non-evolving population, indicating a possible fall in star formation
rate.

This conclusion is supported by recent results from the 8m Subaru
telescope which has now been used for a number of searches for high
redshift Lyman-$\alpha$ emitters using narrow band filter
selection. Taniguchi et al.\ (2003) provide a review of recent
observations of high redshift Lyman-$\alpha$ emitters.  They reproduce
the Lyman-$\alpha$ emitter `Madau' plot from Kodaira et al.\ (2003) and
report a fall by a factor of 4 in the global star formation rate between
$z\sim3$ and $z\sim6.6$, mirroring the decline we from our
Lyman-break-selected $i'$-drop sample.

\subsection{Impact of Evolution of the Luminosity Function or {\em IMF}
on Determining the Global Star Formation Rate} 
\label{sec:lum}

It is fundamental to any discussion of the evolution of basic properties
such as the comoving volume-averaged star formation density that like
must be compared to like. Since our comparison with the work of previous
authors is dependent on assumptions of non-evolution in both the stellar
initial mass function and luminosity function of Lyman break galaxies we
now consider the possible influence on our result of evolution in either
of these functions.

Since the UV luminosity-SFR conversion factor extrapolates the total 
star formation rate given the observed flux from blue, high mass stars,
it is sensitive to the shape of the IMF assumed. In common with previous 
authors we use a Salpeter IMF, however, if the Scalo (1986) IMF is used, 
the inferred star formation rates are a factor of $\approx 2.5$ higher for 
a similar mass range.

Recent work by several authors, (e.g Abel et al.\ 2002, Nakamura \&
Umemura 2002) has suggested that the earliest population of stars in the
universe (population III) has an initial mass function (IMF) that is
biased, with respect to the Salpeter IMF used here, towards massive
stars.  This implies that if such early stellar populations are dominant
at $z\sim6$ then we are likely to be overestimating the star formation rate 
(although this clearly depends on the detailed shape of the IMF).

Recent results (Kogut et al.\ 2003), however, favor an end to reionization 
as early as $z\sim11$ (depending on the reionization history assumed). By
$z\sim6$, therefore, the era of population III stars may well be at an
end. The semi-analytic models of Somerville and Livio (2003) also
suggest that the ionizing flux due to population II stars (assumed to have a
Salpeter IMF) outstrips that of the more massive population III by a
factor of $\sim100$ by $z\sim6$. As a result, we consider that use of a
standard IMF is more appropriate for our calculation at $z\sim6$.

As described in section \ref{sec:phot} our magnitude limit corresponds
to $m^\ast$ for a Lyman-break galaxy at $z\sim6$ if no evolution is
assumed in the Lyman-break galaxy luminosity function found at
$z\sim3$ and applied at $z\sim4$ by Steidel and coworkers (1999). In
figure \ref{fig:madau} we chose to recalculate the reported global
star formation densities of other authors to allow a comparison with
our SFR$>15\,h^{-2}_{70}\,M_{\odot}\,{\rm yr}^{-1}$ limit and observe
a significant fall in star formation density when making this
assumption (although an assumed luminosity fucntion is necessary in
order to calculate the effective volume of the survey). This apparent
fall may also arise if the luminosity function evolves
significantly in this redshift range (a period between 0.9 and 1.5 Gyr
after the Big Bang).

The galaxy
luminosity function $\Phi({\mathrm L})$ is conventionally parameterized
by a Schecter (1976) function of the form:
\begin{displaymath}
\Phi (L)dL=\phi^{*}\exp\left(\frac{\mathrm{-L}}{\mathrm{L}^{*}}\right)\left(\frac{\mathrm{L}}{\mathrm{L}^{*}}\right)^{\alpha}d\left(\frac{\mathrm{L}}{\mathrm{L}^{*}}\right)
\end{displaymath}
Steidel et al.\ (1999) found no conclusive evidence for evolution of
the rest-frame ultraviolet luminosity function between $z\sim3$ and
$z\sim4$ deriving a a faint-end slope $\alpha=-1.6$, a typical
magnitude $m^{\ast}=24.48$ and a normalization of $\phi^\ast=3.5
\times 10^{-2} ,h^{3}_{70}\,{\rm Mpc\ }^{-3}$ (in our cosmology) at $z\sim3$. If we
assume no evolution in $\alpha$ or L$^\ast$ then our result
corresponds to a $\phi^{\ast}_{z=6}=0.1 \phi^{\ast}_{z=3}$.

Figure \ref{fig:lum} illustrates the locus of parameter values that
would produce our results if we instead assume that the comoving star
formation density remains unchanged between $z\sim3$ and $z\sim6$ . 
Also shown are locii corresponding to a moderate fall in the star formation 
density and for a fall as shown in figure \ref{fig:madau}.

\begin{figure}
\plotone{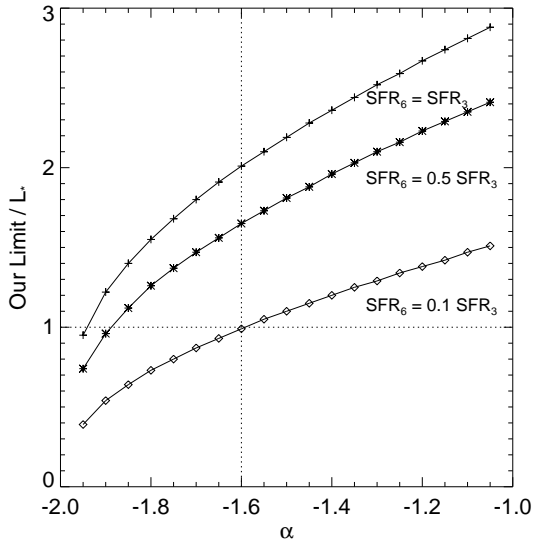}
\caption{Evolution of the Luminosity Function. In the event that there
is no fall in the comoving volume averaged star formation density
between $z\sim3$ and $z\sim6$ our apparent fall could be explained by
a variation in the parameter of the galaxy luminosity function as
shown here (crosses). Also shown are the case of a moderate fall in
the star formation density (asterisks) and for a fall in the
luminosity function as shown in figure \ref{fig:madau}
(diamonds). Dotted lines indicate the parameter values found at
$z\sim3$ by Steidel et al.\ (1999).}
\label{fig:lum}
\end{figure}

As can be seen, if the star formation density is in fact unchanged
from that at $z\sim3$, then our results may be explained either by a
fall in the characteristic luminosity of Lyman break galaxies at
$z\sim6$ together with a corresponding rise in object number density
(L$^{\ast}_{z=6}=0.5$L$^{\ast}_{z=3}$, $\alpha=-1.6$,
$\phi^{\ast}_{z=6}=2\phi^{\ast}_{z=3}$) or by a steepening of the
faint end slope ($\alpha=-2.0$, L$^{\ast}_{z=6}=$L$^{\ast}_{z=3}$,
$\phi^{\ast}_{z=6}=\phi^{\ast}_{z=3}$) or by some combination of the
two.  In practise, either extreme falls outside the expected range of
values for these parameters at $z\sim6$.

\subsection{Comparison with Previous Studies} 
\label{sec:previous_studies}

In recent months a number of groups have made studies of Lyman break
galaxies and star formation histories at $z\ge5$ utilizing deep space or
ground based imaging. In this section we compare these studies and
contrast their results with those presented above.

Multi-wavelength ground-based imaging and the photometric redshifts 
that may be derived from it are valuable tools in the identification
of high redshift objects but are limited by the resolution and 
magnitude limits that can be reached. Fontana et al.\ (2002) presented 
a sample of $z>4.5$ candidates selected at $z'_{\rm AB}<25.0$ from 
imaging with the Very Large Telescope (VLT). They find 13 high redshift
candidates including 4 in their highest redshift bin of $5<z<6$ in a
total area of 29.9 arcmin$^2$ and observe an order of magnitude
decrease in the UV density between $z\sim4.5$ and $z\sim6$. 
However, correcting for their selection function and absolute magnitude limit 
in each redshift bin they report no decrease in the comoving global star
formation rate. This is a relatively small survey and the authors note
significant variance between their two survey fields. In figure
\ref{fig:madau}, we show star formation densities derived using our
prescription from the UV luminosity densities reported by Fontana et
al.\ for their magnitude-limited `minimal' sample of objects which they
are confident lie at high redshift. As can be seen their results are in
good agreement with our own although based on very small numbers of
objects.

HST imaging such as that studied in this paper allows such surveys to be 
pushed to fainter and more distant objects.  At $z>4.5$
Lehnert and Bremer (2003) presented a sample of objects selected on the 
basis of their large $(R-I)_{\rm AB}$ color. Their
number counts at $I_{\rm AB}<26.2$ are slightly lower than expected if
no assumption in the luminosity function is assumed but not
significantly so given the effects of completeness corrections and
cosmic variance. Similarly Iwata and coworkers (2003) consider a catalog
of candidate $z\sim6$ objects selected primarily on the basis of large
$V-I$ colors. The star formation density estimated by Iwata et al.\
from their survey indicates a slight fall in global star formation
density in the range $4<z<5$ although their result could also be interpreted 
as consistent with a constant star formation density at $z>2$. Capak et al.\ 
(2003) comment on the selection criteria in both these papers and suggest that 
both groups may have underestimated the contamination in their sample due to
interloper galaxies at lower redshifts. If this is the case, then the
star formation density derived by Iwata et al.\ for $z\sim5$ (shown in
figure \ref{fig:madau}) may in fact represent an overestimate. This
would support our observations of a fall in global star formation
density with redshift.

Yan et al.\ (2002) also used deep HST imaging to identify 30 $z\sim6$
candidates in an area of 10 armin$^2$ to a depth of $z'_{\rm AB}=28.3$
and with a median magnitude of $z'_{\rm AB}=27.4$. They estimate their
possible contamination from cool dwarf stars and elliptical galaxies as
7 objects (23 per cent), a lower fraction than that estimated by most
other authors. None of the objects identified by these authors would be
selected at our limiting magnitude, consistent with our expected
$0.2\pm0.1$ objects in a 10 arcmin$^2$ field. In common with Bouwens et
al.\ (2003a), we note the difficulty of identifying objects so faint
with certainty given the total exposure times quoted by Yan et
al. Nonetheless, we note with interest the suggestion made by these
authors that their number counts at faint magnitude suggest that the
faint end slope, $\alpha$, of the galaxy luminosity may be as steep as
$\alpha=-2$ (see section \ref{sec:lum} for further discussion of this).

Finally a selection of $z\sim6$ candidate objects made from two deep
fields observed with the {\em HST/ACS} instrument has been presented by
Bouwens et al.\ (2003a) They find a total of 23 objects to a magnitude
limit of $z'_{\rm AB}=27.3 (6\sigma)$, somewhat fainter than our limit,
and use their data (selected with a similar $(i'-z')_{\rm AB}>1.5$ cut
to the sample in this paper) to make three estimates of the star
formation density at this redshift.  The third estimate, presented by
Bouwens and coworkers as a strict lower limit, is largely consistent
with the results presented in this paper (given the different survey
depth) and is derived using a similar methodology and assumption of no
evolution in the intrinsic characteristics of Lyman break galaxies
between $z\sim3$ and $z\sim6$. However, it is notable that their sample
includes only two objects that would be selected in this paper, broadly
consistent with the $1.1\pm0.5$ objects we would expect to observe in
their 46 arcmin$^2$ survey area. Their sample also includes one very
luminous object which substantially increases their final result. In
addition Bouwens et al.\ present two incompleteness-corrected estimates
which make use of the `cloning' procedure described in Bouwens et al.\
(2003b).  This procedure projects a sample of V-drop selected objects
(selected to lie at $z\sim5$) to $z\sim6$ taking into account the
effects of k-corrections, surface brightness dimming and PSF
variations. This sample is then used to assess their survey
incompleteness and selection function, leading to star formation
densities some 2.7-3.8 times higher than their third estimate. This
procedure takes into account the natural variability in size, shape and
surface brightness of galaxies but in doing so makes the implicit
assumption that galaxies at $z\sim6$ have similar distributions in these
properties to galaxies at $z\sim5$. In addition, their first (higher)
estimate is based on comparison with their earlier work at lower
redshifts (Bouwens et al.\ 2003b) which is based on a relatively small
observation area which substantially influences their final result.

The observed trend in the global star formation density at $z\ge3$ is
still uncertain and the evidence mixed. Although this work suggests a
fall in star formation density based on the bright end of the
luminosity function, other, deeper, surveys suggest that it is
possible that there is no fall in the integrated star formation rate.
This suggests that the shape of the luminosity function may well be
evolving over this redshift interval, as discussed in section
\ref{sec:lum}. The results of Bouwens and coworkers in particular,
however, illustrate the complexities involved in fully assessing the
completeness and selection function of a photometrically derived
galaxy sample at $z\sim6$ and the necessity of making any completeness
corrections assumed completely transparent in order to allow
comparison of work in this field.

\subsection{Comparison with the results of the GOODS team}

During the refereeing process of this manuscript a number of papers
written by the GOODS team (described in Giavalisco et al.\ 2003a), and
based on three co-added epochs of GOODS data, have become public.

A colour selection similar to that in this paper and that of Bouwens
et al.\ (2003) was performed by Dickinson et al.\ (2003) in order to
identify $z\approx6$ candidates with a slightly bluer colour cut
criteria of $(i'-z')_{\rm AB}>1.3$ which may leave them susceptible to
increased contamination from lower redshift galaxies. Rather than
apply a strict magnitude limit, these authors have chosen to apply a
cut based on object signal-to-noise in their co-added images.  They
have also limited their survey area to that observed in all three
epochs (approximately 80\% of the total area) and limited their sample
to resolved objects. As a result, a large fraction of the $z\sim6$
candidates described in this paper and in paper 1 do not form part of
their survey. Dickinson et al.\ identify only five candidate objects
with signal-to-noise greater than 10 in the combined GOODS-N and
GOODS-S survey of which three objects (in the GOODS-S field) were also
selected in paper 1 and the remaining GOODS-S object falls outside our
colour and magnitude cuts. The remaining candidate, NiD001, has
colours which appear to vary significantly between epochs and averages
across 4 epochs as too faint to appear in our selection. These authors
also present a deeper sample of galaxies with signal to noise $>5$
although they acknowledge that this is likely to contain significant
contamination. They note that the observed number density of bright
galaxies is significantly lower than that expected from the lower
redshift luminosity functions for Lyman break galaxies, supporting our
conclusion that the luminosity function of these objects must evolve
(either in shape or normalisation) between $z\sim3$ and $z\sim6$.

Giavalisco et al.\ (2003b) utilise the Dickinson et al.\ candidate
selection at $z\sim6$, together with similar colour selections at
lower redshift, to comment on the evolution of the rest-frame UV
luminosity density (and hence star formation rate) with
redshift. These authors have attempted to correct the low
signal-to-noise Dickinson et al.\ sample for contamination and
incompleteness and work to a faint magnitude limit. This yields a star
formation history that declines only slowly with redshift if evolution
in the luminosity function is modelled using a $\chi^2$ approach
(which assumes \textit{a priori} $\alpha=-1.6$) and a fall if the
$V_{eff}$ method using the $z=3$ luminosity function is adopted. While
the $\chi^2$ approach has the advantage of simultaneously fitting the
luminosity function and allowing an estimate of the sample volume, and
dows not reuire an assumed lumosity function, it is not clear that the
number counts of $z\sim6$ candidates are sufficient to support this
approach (Dickinson et al.\ only report five objects with confidence).
However, as Giavalisco et al.\ comment, their result is largely in
agreement with that presented here and in paper 1, given the observed
decline in number density of bright galaxies at $z\sim6$.  Their
result of an almost constant luminosity density at $z>3.5$ is highly
dependent on low-significance faint end counts and on the corrections
applied to these. It seems possible that the number counts of faint
objects will indeed show evolution in the shape of the luminosity
function between $z=3$ and $z=6$, yielding a flat star formation
history at high redshift but the data are not currently available to
determine this. As Giavalisco et al.\ comment, the luminosity density
measurement at $z\sim6$ clearly needs to be revisited with deeper
data, which may be possible with the forthcoming release of the HST
Ultra Deep Field (UDF).

\section{Conclusions}
\label{sec:conclusion}

In this paper we have presented a selection of objects in the Hubble
Deep Field North with extreme $i'-z'$ colors and identify a number of
good $z\sim6$ candidates. From these we have calculated an estimate of
the global star formation density at this redshift. Our main conclusions
can be summarized as follows:

\begin{enumerate}
\item 
The surface density of $i'$-drops with $(i'-z')_{AB}>1.5$ and
$z'_{AB}<25.6$ in the GOODS-N is 0.03 objects/arcmin$^2$, similar to that in
GOODS-S reported in paper~1.  This indicates that cosmic variance is not
the dominant uncertainty in measuring the star formation density at
$z\sim 6$ from the GOODS {\em HST/ACS} images although it is possible we 
may have the first suggestion of a group at $z=5.8$ in the GOODS-S field.

\item
We measure the star formation rate from the rest-frame 1500\,\AA\ flux
in $z\sim 6$ galaxies with observed star formation rates
$>15\,h^{-2}_{70}\,M_{\odot}\,{\rm yr}^{-1}$ (uncorrected for 
extinction).  We detect a decline in the comoving volume-averaged star
formation rate between $z\sim 3$ to $z\sim 6$ for resolved objects at
this bright limit, consistent with the results reported in
paper~1. The observed fall if all $i$-drops (including unresolved
objects) in the field are considered is less pronounced but this is
likely to signifcantly overestimate the star formation density at
$z\sim6$.

\item
We have considered the possible effect of evolution in the shape and
normalization of the luminosity function; if the total comoving star
formation density (including galaxies well below our flux limit) is in
fact the same as at $z\sim 3,$ then the luminosity function must
change significantly (from $z=3$ to $z=6$, $L^{*}$ drops by a factor
of $2$ if $\phi^{*}$ and $\alpha$ are constant, or alpha changes from
$-1.6$ to $-2.0$ if $\phi^{*}$ and $L^{*}$ remain constant at $z>3$).

\item
Ultra-deep Keck/DEIMOS spectroscopy has been obtained for half our
sample. The long-slit spectrum of GOODS-N object 4 shows a single
emission line at 8807\,\AA , offset by 2\arcsec\ , although it is
possible that this is associated with a nearby galaxy at lower redshift
(not an $i'$-drop). If it is spatially-extended Ly-$\alpha$ emission
then the redshift would be $z=6.24$, although a lower-redshift
interpretation is possible. We detect continua from two sources, with
low signal-to-noise, but these may be consistent with late-type stars.
Two fainter sources have no detectable flux in the spectroscopy.

\item
For the total sample (GOODS N$+$S), we confirm very few $i'$-drop
galaxies in Lyman-$\alpha$ emission (2 or 3 out of $\sim 15$).
This could be that the contamination of low-$z$ galaxies and
stars is significant, or perhaps that Lyman-$\alpha$ does
not routinely escape from $z\sim 6$ galaxies.

\item
We have analyzed surface brightness selection
effects: although undoubtedly these play a role,
in fact the vast majority of our candidates are
very compact (as would be expected for star-forming
regions), so this seems not to be a significant effect.
We correct our survey volume for the dimming in apparent $z'$-band 
magnitude with redshift by using a $V/V_{\rm max}$ ``effective volume''
approach, introducing a correction of about a factor
of two.

\end{enumerate}

\subsection*{ACKNOWLEDGMENTS}

TT acknowledges support from NASA through Hubble Fellowship grant
HF-01167.01. The analysis pipeline used to reduce the DEIMOS data was
developed at UC Berkeley with support from NSF grant AST-0071048, and
we are grateful to Douglas Finkbeiner, Alison Coil and Michael Cooper
for their assistance with this software.  We thank Greg Wirth, Chris
Willmer and the Keck GOODS team for kindly providing the
self-consistent astrometric solution for the whole GOODS field we used
to design the slit-masks. We thank Mark Sullivan for assistance in
obtaining the spectroscopic observations of the GOODS-N and Dave
Alexander, Simon Hodgkin and Rob Sharp for helpful discussions.  We
also thank for anonyomous referee for his useful comments. Some of the
data presented herein were obtained at the W. M.\ Keck Observatory,
which is operated as a scientific partnership among the California
Institute of Technology, the University of California and the National
Aeronautics and Space Administration.  The Observatory was made
possible by the generous financial support of the W. M.\ Keck
Foundation, and we acknowledge the significant cultural role that the
summit of Mauna Kea plays within the indigenous Hawaiian community. We
are fortunate to have the opportunity to conduct observations from
this mountain. This paper is based on observations made with the
NASA/ESA Hubble Space Telescope, obtained from the Data Archive at the
Space Telescope Science Institute, which is operated by the
Association of Universities for Research in Astronomy, Inc., under
NASA contract NAS 5-26555. The {\em HST/ACS} observations are
associated with proposals \#9425\,\&\,9583 (the GOODS public imaging
survey). We are grateful to the GOODS team for making their reduced
images public -- a very useful resource.

\end{document}